\documentclass[11pt,twocolumn]{aastex63}
\usepackage{chngcntr}

\newcommand{\ms}{$\rm m\,s^{-1}$ }
\newcommand{\kms}{$\rm km\,s^{-1}$ }
\newcommand{\msnospace}{$\rm m\,s^{-1}$}
\newcommand{\kmsnospace}{$\rm km\,s^{-1}$}
\newcommand{\vsini}{\textit{v}\,\textnormal{sin}\,\textit{i} }
\newcommand{\logg}{log\,$g$ }

\received{2021 February 1}
\revised{2021 March 9}
\accepted{2021 April 5}
\submitjournal{The Astronomical Journal}

\graphicspath{{./}{figures/}}

\shorttitle{Precision RV with \texttt{IGRINS RV}}
\shortauthors{Stahl et al.}

\begin{document}

\title{\texttt{IGRINS RV}: A Precision RV Pipeline for IGRINS \\
Using Modified Forward-Modeling in the Near-Infrared \footnote{\url{https://github.com/shihyuntang/igrins_rv}}}


\author[0000-0002-0848-6960]{Asa G. Stahl}
    \affiliation{Department of Physics and Astronomy, Rice University, 6100 Main Street, Houston, TX 77005, USA}

\author[0000-0003-4247-1401]{Shih-Yun Tang}
    \affiliation{Lowell Observatory, 1400 West Mars Hill Road, Flagstaff, AZ 86001, USA}
    \affiliation{Department of Astronomy and Planetary Sciences, Northern Arizona University, Flagstaff, AZ 86011, USA}

\author[0000-0002-8828-6386]{Christopher M. Johns-Krull}
    \affiliation{Department of Physics and Astronomy, Rice University, 6100 Main Street, Houston, TX 77005, USA}

\author[0000-0001-7998-226X]{L. Prato}
    \affiliation{Lowell Observatory, 1400 West Mars Hill Road, Flagstaff, AZ 86001, USA}
    \affiliation{Department of Astronomy and Planetary Sciences, Northern Arizona University, Flagstaff, AZ 86011, USA}

\author[0000-0003-4450-0368]{Joe Llama}
    \affiliation{Lowell Observatory, 1400 West Mars Hill Road, Flagstaff, AZ 86001, USA}

\author[0000-0001-7875-6391]{Gregory N. Mace}
    \affiliation{Department of Astronomy, University of Texas at Austin, 2515 Speedway, Austin, TX, USA}
    
\author{Jae Joon Lee}
    \affiliation{Korea Astronomy and Space Science Institute, 776 Daedeokdae-ro, Yuseong-gu, Daejeon 34055, Republic of Korea}

\author{Heeyoung Oh}
    \affiliation{Korea Astronomy and Space Science Institute, 776 Daedeokdae-ro, Yuseong-gu, Daejeon 34055, Republic of Korea}
    
\author{Jessica Luna}
    \affiliation{Department of Astronomy, University of Texas at Austin, 2515 Speedway, Austin, TX, USA}

\author{Daniel T. Jaffe}
    \affiliation{Department of Astronomy, University of Texas at Austin, 2515 Speedway, Austin, TX, USA}

\email{Asa.Stahl@rice.edu, sytang@lowell.edu}

\begin{abstract}

Application of the radial velocity (RV) technique in the near infrared is valuable because of the diminished impact of stellar activity at longer wavelengths, making it particularly advantageous for the study of late-type stars but also for solar-type objects. In this paper, we present the \texttt{IGRINS RV} open source \texttt{python} pipeline for computing infrared RV measurements from reduced spectra taken with IGRINS, a R $\equiv \lambda/\Delta \lambda \sim$ 45,000 spectrograph with simultaneous coverage of the H band (1.49--1.80 $\mu$m) and K band (1.96--2.46 $\mu$m). Using a modified forward modeling technique, we construct high resolution telluric templates from A0 standard observations on a nightly basis to provide a source of common-path wavelength calibration while mitigating the need to mask or correct for telluric absorption. Telluric standard observations are also used to model the variations in instrumental resolution across the detector, including a yearlong period when the K band was defocused. Without any additional instrument hardware, such as a gas cell or laser frequency comb, we are able to achieve precisions of 26.8 \ms in the K band and 31.1 \ms in the H band for narrow-line hosts. These precisions are empirically determined by a monitoring campaign of two RV standard stars as well as the successful retrieval of planet-induced RV signals for both HD\,189733 and $\tau$\,Boo\,A; furthermore, our results affirm the presence of the Rossiter-McLaughlin effect for HD\,189733. The \texttt{IGRINS RV} pipeline extends another important science capability to IGRINS, with publicly available software designed for widespread use. 

\end{abstract}




\section{Introduction} \label{sec:intro}

The radial velocity (RV) method is a powerful tool for detecting planets around other stars. Ever since the analysis of RV variations facilitated the first detection of an exoplanet around a Sun-like star \citep[51 Pegasi\,b, ][]{mayo95}, the technique has remained one of the major workhorses for exoplanet discovery and characterization \citep{akes13}. RV analysis is the only non-serendipitous exoplanet detection method that provides an independent determination of a planet's mass, an essential parameter for planet characterization. RV variations can also directly measure a planet's orbital eccentricity. This is often studied in connection with dynamical evolution and planet formation pathways \citep{mori08,wang17}.

In concert with transit observations, RV measurements make it possible to estimate the bulk densities of planets \citep{marc14} and to measure spin-orbit angles through the Rossiter-McLaughlin effect \citep{coll10,tria18}. At the same time, planet-induced RV signals are less dependent on orbital period and inclination than transit signals \citep{mart19}. RV surveys are thus important for probing the range of potential planetary system architectures.

RVs measured in the infrared, as opposed to the optical, extend the capabilities of the method. For one, they provide a better means for planet searches around M dwarfs, which emit the most light in the near infrared (NIR) \citep{plav15}. These stars are increasingly attractive targets for exoplanet detection efforts on account of their abundance in the Milky Way \citep{henr06}, the increased detectability of small planets around small stars \citep{gunt20}, their closer, more easily probed habitable zones \citep{fran16}, and their amenability to planetary atmospheric characterization \citep{ball19}. 

Infrared RVs also exhibit diminished wavelength-dependent stellar jitter, such as those induced by starspots \citep{robe20,tran21}. As the radiation of both a star and starspot generally follow a Planck distribution, at wavelengths greater than that of their peak emission, the contrast between a spot and its host star will diminish. Studies confirming this effect \citep{huel08,mahm11b,bail12} have found RV spot modulation lowers by a factor of $\sim$3 or more when moving from the optical to the H and K bands. Because planet-induced RV variations are wavelength-independent, comparing measured RVs in the optical and NIR provides a robust means of diagnosing spot-induced false positives.

The capacity of NIR RVs to identify and/or diminish the relative magnitude of RV variations caused by stellar processes means that they are useful for detecting planets around spotted FGKM stars and, notably, young stars. Pre-main sequence stars within the mass regime of the Sun, known as T Tauri stars, are highly active, and their spotted surfaces and variable accretion can easily drown out otherwise observable planetary signals \citep{croc12}. The strong magnetic activity of T Tauri stars \citep{john07a} can drive clumpy accretion and generate surface spots that obscure or mimic planetary signals, for example through apparent RV fluctuations with semiamplitudes on the order of \kms at optical wavelengths \citep{huer08,dumu14}. Even optical signals as large as that of a 20 $\rm M_{Jup}$ companion orbiting 0.2 AU from a 0.6 $\rm M_{\odot}$ host star (a semiamplitude of 1 \kms) can be overwhelmed by typical T Tauri activity \citep[e.g., ][]{prat08}. 

Several planet candidates around young stars have been detected over the last decade \citep[e.g.,][]{vane12, krau14, sall15, dona16, oelk16}. The first planet around a T Tauri star young enough to still host its disk was confirmed last year via direct detection of CO in the planet \citep{flag19}. Additional discoveries of planets around young stars are expected to help clarify the nature of planet formation \citep{livi18,vand18}, migration \citep{daws18}, and photoevaporative atmospheric loss \citep{davi19b} through their direct probing of systems while such processes are ongoing. Over the last several years, several spectrographs dedicated to planet detection in the infrared have come online, including HPF \citep{maha12}, IRD \citep{kota14}, and CARMENES \citep{baue20}. New infrared spectrographs like iSHELL \citep{rayn16}, while not designed for precise RV measurements, are also demonstrating sufficient RV precision to detect short-period and massive planets \citep[e.g., ][]{plav20}, continuing the progress made with legacy instruments such as CSHELL \citep[e.g., ][]{croc12} and CRIRES \citep[e.g., ][]{bean10}.

In this paper, we present \texttt{IGRINS RV}: a precision RV analysis pipeline for the Immersion GRating INfrared Spectrometer \citep[IGRINS, ][]{yuk10,park14,mace16,mace18}. IGRINS is a compact and mobile IR spectrometer that has enabled a variety of science on 3 different telescopes so far. Its large spectral grasp and high resolution facilitate a vast array of applications, from exoplanet detection \citep{mann16} and atmospheric characterization \citep{llam19}, to studies of stars \citep{sterl16}, the ISM \citep{kapl17}, and galaxies \citep{guer19}.

Only a few cases of RV analysis have been performed with IGRINS so far. Using a crosscorrelation technique, \citet{carl18} report typical external RV uncertainties of 59 \msnospace, but they do not monitor an RV standard star to characterize the internal precision of their method, thus their total, longterm RV precision is unknown. Another published analysis employing crosscorrelation achieves robust median uncertainties of 150 \ms \citep{mace16b}. Lastly, \citet{john16} applied a forward modeling technique using a static telluric template to achieve a precision of 75 \msnospace. \texttt{IGRINS RV} represents a more sophisticated continuation of this forward-modeling methodology, whose previous versions have also been effectively applied to CSHELL and PHOENIX spectra in the past \citep{croc12}. The code will allow anyone to process IGRINS data to final RVs with nearly 3 times better precision than these previous studies, in an entirely self-contained package easily run from the command line. 

IGRINS is a powerful instrument with a wide variety of science capabilities. At the same time, it is not temperature controlled at the level required for $\sim$\ms RVs, is not fiber-fed, and does not have a built-in gas cell, laser frequency comb, or Fabry-Perot etalons---in short, it was not designed for precise RV experiments. \texttt{IGRINS RV} calculates precise NIR RVs through a modified forward modeling technique in which a data-driven synthetic telluric template leverages atmospheric absorption lines as a wavelength calibrator. The methodology by which it does this is novel and could potentially be applied to other spectrographs. 

The structure of this paper is as follows: in Section~\ref{sec:data}, we describe the data used in testing and validating the pipeline; in Section~\ref{sec:pipe}, we outline how the pipeline works; in Section~\ref{sec:results}, we present the results of applying our code to RV standard stars and known planet hosts; in Section~\ref{sec:conc}, we summarize our findings and describe their future prospects.

Throughout this paper, we refer to each spectrum taken in a nodding sequence (e.g., A) as an ``observation", and the collection of all observations in a sequence (e.g., AB or ABBA) as an ``exposure".


\section{Observations and Data Reduction} \label{sec:data}

IGRINS is a high resolution (R$\equiv \lambda/\Delta \lambda \sim$45,000) cross-dispersed echelle spectrograph with a broad spectral grasp covering the full H band (1.49--1.80 $\mu m$) and K band (1.96--2.46 $\mu m$) simultaneously \citep{yuk10,park14,mace16,mace18}. The former is split into 28 orders and the latter is split into 26 orders, where each order corresponds to $\sim 240$ $\rm \AA$ and consists of 2048 pixels. The H and K bands are imaged on two separate detectors.


IGRINS saw first light on McDonald Observatory's 2.7 m Harlan J. Smith Telescope in 2014. It 
was then moved to the Lowell Discovery Telescope (LDT, formerly the Discovery Channel  Telescope, DCT) in August 2016. Currently, IGRINS is available as a visiting instrument at the Gemini South Telescope. In this paper, we only use data taken while IGRINS was mounted at McDonald Observatory (McD) and at Lowell Observatory. 


Spectra taken at the DCT included information on environmental conditions in their fits headers, such as the ambient humidity and temperature during an observation, but spectra taken at McD lacked these data. This led to a slightly different treatment of the observations when constructing telluric templates.


Observing logs for science targets can be found in Appendix Tables \ref{tab:obs_std} and \ref{tab:obs_tar}. The telluric standards used are listed in Appendix Table \ref{tab:A0s}.

\subsection{IGRINS K band Defocus}

\begin{figure}[tb!]
\centering
\includegraphics[width=0.45\textwidth]{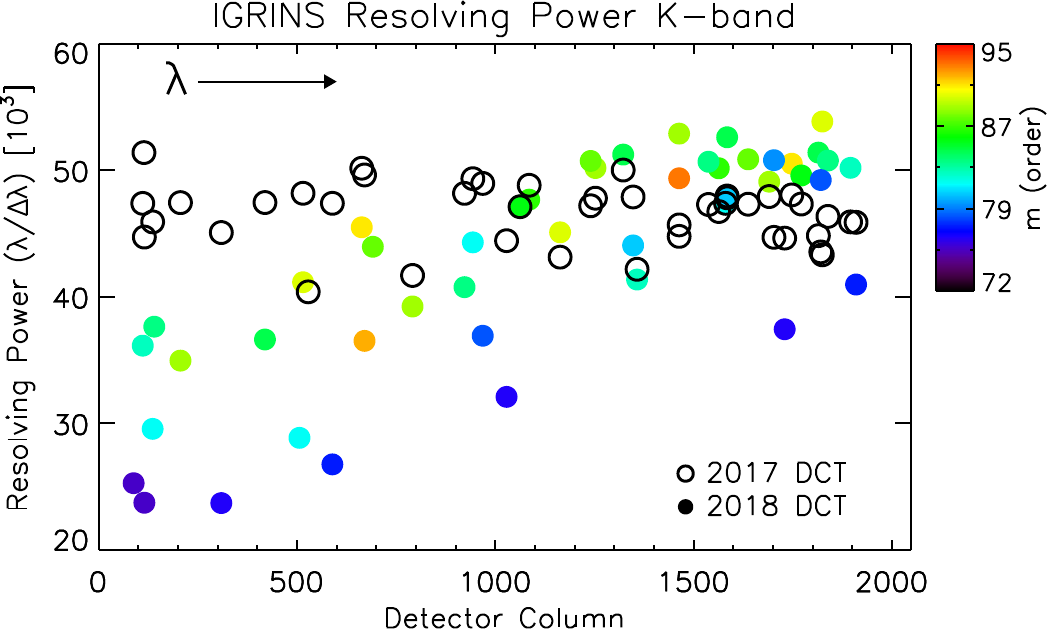}
\caption{
   IGRINS spectral degradation in K band measured from the FWHM of sky OH emission lines, showing the changes in spectral resolution between 2017 and 2018 as a function of detector column and the diffraction order (m) on the detector. 
    }
\label{fig:defocus}
\end{figure}

Between January 2018 and August 2019, IGRINS exhibited a change in spectral resolving power in the K band as the result of loose fasteners on the back of the detector mount. Figure~\ref{fig:defocus} shows how the effect of this defocus varied over the detector. The largest change to the echellogram, in which the spectral resolution was reduced to R$\sim$20,000, occurs in the short-wavelength end of the long-wavelength K band orders (wavelengths on the IGRINS detector increase with increasing detector column and towards lower spectral orders). This is where the CO bandheads are located ($\lambda$ $>$ 2.3 $\mu$m). While only two epochs are shown here, we have verified that these trends persist across many nights.

The defocus affected all data taken during the 2018 visit to Gemini and the subsequent visit to the DCT. In August 2019, all fasteners on the K band detector mount were re-pinned and tightened in the lab at the University of Texas at Austin. Cold testing showed that the echellogram focus was corrected back to the original design specifications \citep{yuk10,park14}. The H band detector was also moved away from its camera by 200 $\mu$m to achieve better focus. Laboratory testing at Gemini South in February 2020, prior to recommissioning and after shipment from UT Austin, confirmed that the IGRINS H and K band echellograms remained at optimal focus.

The strategies \texttt{IGRINS RV} uses to mitigate the impact of the defocus are described in Sections \ref{ssec:ip} and \ref{ssec:err}.

\subsection{Data Reduction and Flux Suppression Effect} \label{ssec:reduction}

We employed the IGRINS pipeline package version 2.2.0 \citep[plp\,v2.2.0\footnote{\url{https://github.com/igrins/plp}};][]{leegul17}
 to reduce our data from a raw format to one-dimensional spectra. While this reduction pipeline generates a wavelength solution, it is not the final wavelength solution used by \texttt{IGRINS RV}, as will be discussed in Section \ref{sec:pipe}.

To facilitate robust uncertainty estimates in our RV measurements, we performed the extraction of the individual A and B frames of our observations separately. Because the A and B observing settings produce different slit illuminations, we were able to accomplish this by instructing the reduction pipeline only to use the upper or lower half of the echellogram orders through its \texttt{--frac-slit} command flag. 

In the process of optimizing the RVs estimated from these individual A and B frames, we noticed a subtle artifact in the data: reduced A frames often exhibited a discrete reduction in flux near the peak of the blaze (Figure \ref{fig:blazedip_fix}~a and b). While the exact cause of this artifact is not known at this time, the flux suppression appears to be the result of a hardware effect that is exacerbated by the way the current version of the reduction pipeline (plp\,v2.2.0) performs optimal extraction. While the cause will be diagnosed and addressed if possible, an upcoming release of the plp will fully correct for the effect. In the meanwhile, it is relatively straightforward to address the flux reduction in our analysis (Section~\ref{ssec:model},  Figure~\ref{fig:blazedip_fix}~c). Once the plp is updated, a new version of \texttt{IGRINS RV} will be correspondingly released. This effect is not present in spectra reduced by plp in telluric-corrected mode.


\subsection{RV Standards: GJ\,281 and HD\,26257} \label{sec:RVstandard}

GJ\,281 and HD\,26257 are both recognized RV standard stars. Past monitoring has found GJ\,281 \citep[M0.0Ve, \vsini = 2.96 \kmsnospace,][]{lepi13} to be non-variable in RV to $<$ 12 \ms \citep{endl03}, and HD\,26257 \citep[G2V, \vsini = 9.24 \kmsnospace,][]{mich99} to be non-variable to $<7$ \ms \citep{butl17}. Additional information on these stars can be found in Table~\ref{tab:info}.

Observations of GJ\,281 took place mainly at DCT between December 2016 and March 2019. Of 60 nights of observation, 22 took place during the IGRINS K band defocus. We also drew on 6 nights of observations taken at McD between November 2014 and November 2015. 

HD\,26257 was observed a total of 34 times at DCT between November 2016 and January 2018.

Both RV standards are relatively bright, so most exposures typically consisted of an AB nodding sequence, as opposed to (e.g.) an ABBA sequence.

To assure the quality of the extracted RVs, we refrained from analyzing observations without an accompanying telluric standard taken within an airmass difference of 0.3, nor did we analyze RV standard spectra with S/N lower than 50. These cuts left us with 60 total observations for GJ\,281, 5 of which were taken at McD, and 33 observations for HD 26257. See Appendix Table~\ref{tab:obs_std} for details.



\subsection{Known Planet Hosts}

In addition to our RV standards, we tested the performance of \texttt{IGRINS RV} by applying it to two known planet-hosting stars, HD\,189733 and $\tau$\,Boo\,A. 


\subsubsection{HD\,189733\,b} 

HD\,189733\,b is a transiting $1.15\,\pm\,0.04$ M$_{J}$ planet in a $\sim$2.219 day orbit around its host star. It was first detected by \citet{bouc05} using the ELODIE fiber-fed optical spectrograph. This planetary system is in a nearly edge-on orientation with an inclination of $85.3\,\pm\,0.1\degr$ and induces an RV semi-amplitude in its host star of $205\,\pm\,6\,$\msnospace. \citet{bako06} reported a distant M-dwarf companion with a projected separation 11.2\arcsec and a period of $\sim$3200 years, but no follow up studies have attempted to refine the stellar companion's orbital solution. Other than being a famous target for planetary atmospheric studies \citep{alon19a,stei19}, HD\,189733\,b is also one of the few planetary systems exhibiting the Rossiter-McLaughlin effect. The system's transit-induced RV anomaly was first recognized by \citet{bouc05} and has since been revisited with NIR observations using SPIRou \citep{mout20}, who found the peak-to-peak amplitude of the anomaly to be $\sim$110 \msnospace. Other basic information on this target can be found in Table~\ref{tab:info}.

Observations of HD\,189733 took place at McD between May 2015 and July 2016 and at DCT between October 2016 and September 2017. All 8 nights of observations taken were used. Given the short period of HD\,189733\,b, each night of observation covered a large portion of its phase space. We therefore treated each exposure (i.e., each ABBA sequence) as a statistically independent RV measurement. Of 82 such measurements, we were able to extract 81 high quality RVs. 



\subsubsection{$\tau$\,Boo\,A\,b} \label{sssec:taub}

$\tau$\,Boo\,A\,b was one of the first exoplanets ever discovered \citep{butl97}. The gas giant orbits in a roughly circular 3.31 day orbit, inducing an RV semi-amplitude of $461.1\,\pm\,7.6\,$\ms \citep{butl06}. Two separate teams have estimated the inclination of the orbit to be $44.5\,\pm\,1.5\degr$ \citep{brog12} and $47\,\pm\,7\degr$ \citep{rodl12}, corresponding to planetary masses of $\rm 5.95\,\pm\,0.28\,M_{J}$ and $\rm5.6\,\pm\,0.7\,M_{J}$, respectively. 

A visual companion M star to $\tau$\,Boo\,A is at a separation of $1.662\pm0.002$\arcsec \citep{jus19}. Several studies have tried to estimate the orbital solution for the $\tau$\,Boo\,AB binary system \citep{hale94,popo96}, but these efforts have been complicated by poor orbital phase coverage as a result of the long period of the system. With over 150 years of astrometric data and 25 years of RV data, \citet{jus19} were unable to place precise constraints on the binary's orbital period or semi-major axis, estimating $2420^{+2587}_{-947}$ years and $221^{+138}_{-62}$ AU, respectively. However, the authors do confirm the eccentricity of the orbit at $0.87^{+0.04}_{-0.03}$, and their RV data provide a means through which one can account for the M dwarf's impact on the measured RVs of $\tau$\,Boo\,A (Section~\ref{ssec:phosts}). Besides affecting the actual velocity of $\tau$\,Boo\,A, we expect light from the M dwarf companion to slightly contaminate IGRINS spectra of the target, given the instrument's slit width of 1\arcsec\ at McDonald, and that the H band magnitude difference between the two stars is $\sim$4 mag based on their blackbody temperatures. Table~\ref{tab:info} provides additional information on the planet host.

The observation of $\tau$\,Boo\,A took place at McD over two epochs, the first lasting from January to July of 2015, and the second between February and March of 2016. As with HD\,189733, the short period of the planetary companion and the intense monitoring cadence led us to divide each night into multiple independently-analyzed ABBA exposures, such that the 16 nights of data yielded 217 RV measurements, of which 215 were of high enough quality to deliver reliable values.



\begin{deluxetable*}{lRR l LL RR C l RR}
\tablecaption{Test Targets Basic Information\label{tab:info}
             }
\tabletypesize{\scriptsize}
\tablehead{
	 \colhead{Target}          &  \colhead{R.A.}        & \colhead{Decl.}           & \colhead{SpTy}        & \colhead{T$_{\rm eff}$}       & \colhead{\logg}           &
	 \colhead{H$^{a}$}         &  \colhead{K$^{a}$}     & \colhead{Distance$^{b}$}  & \colhead{Notes}       & \colhead{RV$^\dagger$}        & \colhead{\vsini$^\dagger$}\\
	 \colhead{ }               & \colhead{(hh:mm:ss)}   & \colhead{($\pm$dd:mm:ss)} & \colhead{ }           & \colhead{(K)}                 & \colhead{ }               &
	 \multicolumn{2}{c}{(mag)}                          & \colhead{(pc)}            & \colhead{ }           & \multicolumn{2}{c}{(\kms)}                                \\
	 \cline{7-8} \cline{11-12}
	 \colhead{(1)} & \colhead{(2)} & \colhead{(3)} & \colhead{(4)} & \colhead{(5)} & \colhead{(6)} & \colhead{(7)} &
	 \colhead{(8)} & \colhead{(9)} & \colhead{(10)} & \colhead{(11)} & \colhead{(12)}
     }
     
\startdata 
GJ\,281          & 07\colon39\colon23.04    & +02\colon11\colon01.2     & M0.0Ve$^{c}$    & 4014$^{g}$  & 4.66$^{g}$ & 6.09 & 5.87  & 15.05 & RV Standard & 20.08    & 2.96\\
HD\,26257        & 04\colon09\colon09.07    & +00\colon10\colon44.3     & G2V$^{d}$       & 6129$^{h}$  & 4.31$^{h}$ & 6.44 & 6.42  & 56.05 & RV Standard & 33.92    & 9.24 \\
HD\,189733       & 20\colon00\colon43.71    & +22\colon42\colon 39.1    & K2V$^{e}$       & 5023$^{h}$  & 4.51$^{h}$ & 5.59 & 5.54  & 19.76 & Planet Host & \nodata  & \nodata \\
$\tau$\,Boo\,A   & 13\colon47\colon15.74    & +17\colon27\colon 24.9    & F7IV-V$^{f}$    & 6387$^{i}$  & 4.27$^{i}$ & 3.55 & 3.36  & 15.65 & Planet Host & \nodata  & \nodata  
\enddata
\tablecomments{$^\dagger$ Determined by \texttt{IGRINS RV} (this study),\\
    $^{a}$ \citet{skru06}, $^{b}$ \citet{bail18},
    $^{c}$ \citet{lepi13}, $^{d}$ \citet{mich99}, $^{e}$ \citet{gray03}, 
    $^{f}$ \citet{gray01}, $^{g}$ \citet{sch19},  $^{h}$ \citet{bre16}, $^{i}$ \citet{fisc05}
    }
\end{deluxetable*} 

\section{The \texttt{IGRINS RV} Pipeline} \label{sec:pipe}



\subsection{Workflow}

The \texttt{IGRINS RV} pipeline is divided into three main steps: Telluric Modelling, Initial Convergence, and Analysis. Each is provided in the package as a separate module and each is run from the command line with keywords specifying all relevant information. A brief outline of each step follows.

\textbf{Step 1 - Telluric Modelling:} Defines the wavelength regions to be analyzed; generates a synthetic, high-resolution telluric template for use in later model fits on a night by night and frame by frame (i.e., A vs B) basis. 

\textbf{Step 2 - Initial Convergence:} Required if the average RV of the target star, or its \vsini, is unknown to $>$ 5 \kms precision. This step performs an abbreviated analysis of the target star observations in order to converge to coarsely accurate RVs, which will be used as starting points for the more precise analysis in the next step. It simultaneously does the same for the target star's \vsini, if unknown. For the sake of expediency, only a single  wavelength region (see Section~\ref{ssec:regions}) is used, and only a single B frame observation for every given exposure is fit. 

\textbf{Step 3 - Analysis:} Performs a full analysis of each target star observation to produce accurate and precise RVs. All the wavelength regions defined in Step 1 are used, and the code performs spectral fits of each observation that is part of a given exposure separately. 

Unless the target \vsini is already known to high accuracy, an initial run of Step 3 in which \vsini is allowed to vary is required. This provides an estimate of \vsini that can then be plugged into the code as a fixed value in the second run of Step 3. If \vsini is already well-known, it is not necessary to run Step 3 more than once, as the code fully converges to the final RVs (within uncertainty) through just one run.


\subsection{Stellar Template Generation} \label{ssec:stell}


The stellar templates used in the analyses presented in this paper were primarily produced with the SYNTHMAG C++ code \citep{koch07} in concert with PHOENIX NextGen model atmospheres \citep{haus99} and the VALD stellar line database \citep{ryab15}. VALD line lists were generated with a detection threshold of 0.01, a microturbulence of 1 \kmsnospace, and solar chemical composition. The effective temperature and \logg were set to within 200 K and 0.5 dex step sizes, respectively, to those of the target star in question.

For GJ\,281, we used a synthetic stellar template with a T$_{\rm eff}$ of 4000\,K and \logg of 4.5; for HD\,26257, we used a template with a T$_{\rm eff}$ of 6200\,K and \logg of 4.5. The HD\,189733 analysis employed a template with a T$_{\rm eff}$ of 5000\,K and \logg of 4.5. For $\tau$\,Boo\,A, we used a template with a T$_{\rm eff}$ of 6400\,K and \logg of 4.5.

It is not uncommon for RV experiments to construct a stellar template from observations \citep[e.g., ][]{cale18}. However, extraction of a stellar template from data requires computationally expensive iteration and disfavors the use of separate exposures for the sake of a combined, higher S/N spectrum. Our ability to analyze separate exposures in a nodding sequence provides better characterization of our precision.

All templates used in this study are provided with the \texttt{IGRINS RV} package. This includes the four templates mentioned above, as well as several others used for testing the code. The models unevenly span a temperature range of 3000--6400 K and \logg of 3.5--5.0. Users running the code on any target stars beyond the range of effective temperature or surface gravity of the handful of templates provided are strongly encouraged to supply their own stellar templates, as discussed in Section~\ref{ssec:disc_templates}.


\subsection{Telluric Template Generation} \label{ssec:tell}

In the PHOENIX \citep{hink98} and CSHELL \citep{gree93} RV codes deployed in \citealt{john16}, the telluric contribution to spectral absorption is modelled using the high-resolution (R $\sim$ 600,000) atlas of the infrared solar spectrum taken with the Fourier Transform Spectrometer at the McMath/Pierce Solar Telescope on Kitt Peak \citep{livi91}. \citet{livi91} monitored the solar infrared spectrum over time and used the stellar absorption's variability to carefully distinguish it from telluric absorption. This isolated telluric spectrum is hereafter referred to as the ``Livingston atlas". 

The Livingston atlas was adequate when applied over the small wavelength ranges covered by CSHELL and PHOENIX ($\sim$60 and $\sim$115 $\rm\AA$, respectively), but when applied across several orders of IGRINS spectra, it results in large discrepancies between the model and the data, jeopardizing the accuracy of the model fits.

Pairing target observations with those of bright, featureless stars (typically type A0) can ameliorate this issue. One method is to use the A0 spectra themselves as telluric templates. Observed soon before or after the target star and at similar airmasses and sky location, such A0 templates will have very similar telluric contributions as the target star spectra, and will also exhibit comparable instrumental broadening. However, this strategy suffers from two major drawbacks: the A0 spectra are sampled at the same resolution as the target star spectra, such that the user's choice to interpolate the observed spectrum as a template will cause over-interpolation, and, as observations in themselves with practical limitations on their S/N, they insert additional noise into the fitting process. 

We employ a strategy intended to optimize the benefits of A0 telluric templates without any of the drawbacks (except the additional observing time required). Using the \texttt{Telfit} \citep{gull14} package, which is itself a \texttt{python} implementation of the \texttt{FORTRAN} Line-By-Line Radiative Transfer Model code \citep{clou05}, we fit the A0 spectrum associated with each target star observation to produce a high-resolution synthetic template from the best fit parameters. Such forward-modeling has been found to outperform other telluric mitigation methods, such as correction and cross-correlation \citep{lato20}. 

\begin{figure}[tb!]
\centering
\includegraphics[angle=0, width=1.\columnwidth]{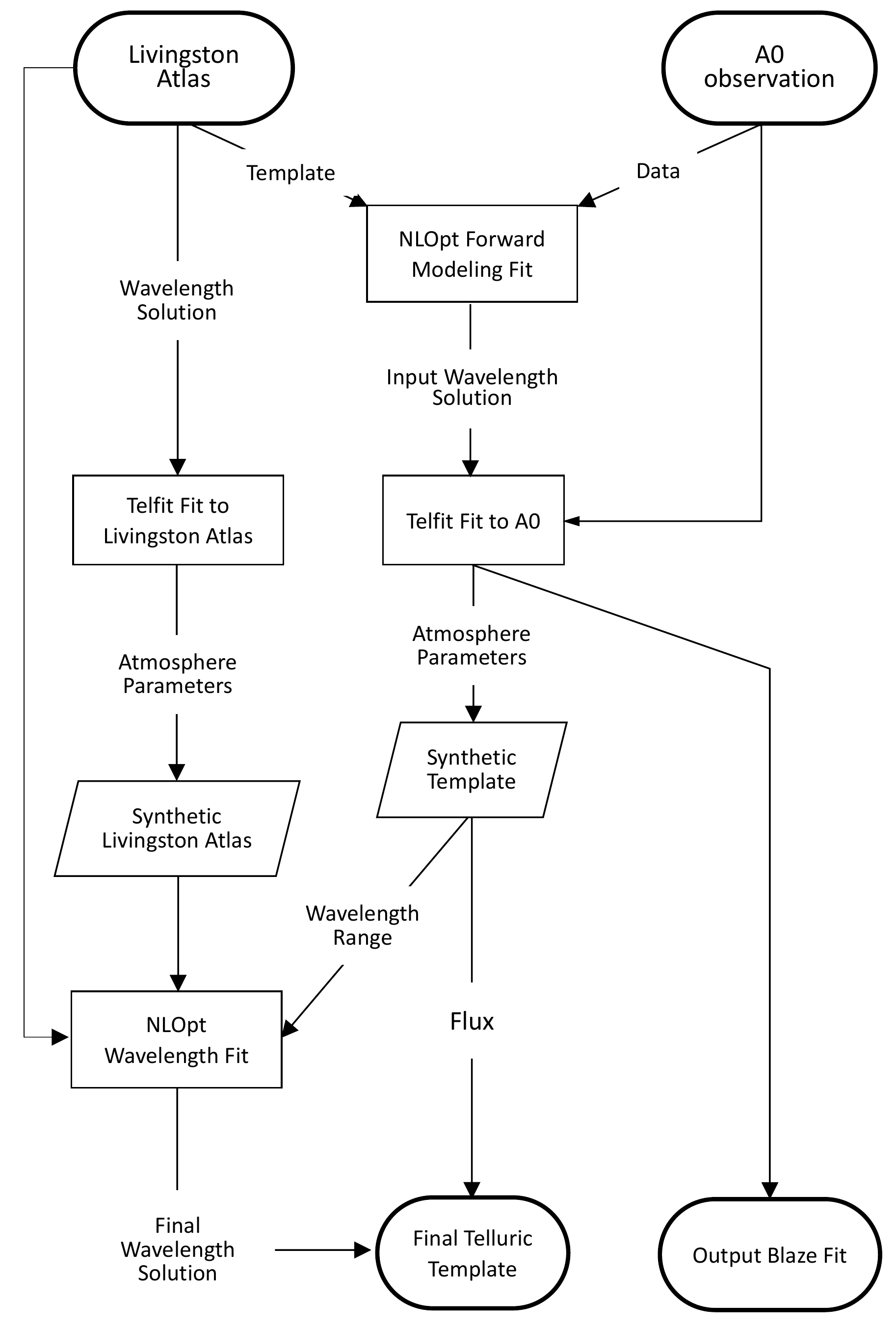}
\caption{
         A schematic outlining the construction of the synthetic telluric template from the Livingston atlas and the A0 observation associated with the target data. The middle column follows the process by which the initial version of the template is produced, while the left column describes the preparation of the wavelength solution it will be combined with to produce the final template. The right column shows how the blaze fit is obtained and subsequently refined. Processes are panelled as squares, internal data products as rhomboids, and inputs/outputs as ovals. 
	    }
\label{fig:flow}
\end{figure}

The synthetic telluric template's accuracy is essential to the performance of \texttt{IGRINS RV}, as it determines the reliability of the model wavelength scale and is a major factor in how well the model fits the data. The pipeline constructs the synthetic template through a series of steps, which are schematically depicted in Figure~\ref{fig:flow}. The entire procedure occurs on an order by order basis for all specified wavelength regions and is repeated for each telluric standard associated with the target star observations. Brackett series stellar absorption does not contaminate the process, as those lines happen to coincide with IGRINS orders that have too little telluric absorption to be used by our pipeline (Section~\ref{ssec:regions}).

First, the pipeline builds an input wavelength solution associated with the A0 spectrum that will be plugged into \texttt{Telfit}. This solution is provided by applying our standard \texttt{NLOpt} forward-modeling fit to the A0 observation, employing the Livingston atlas as a telluric template. While the Livingston atlas does not match the observed data well enough to facilitate the measurement of highly precise RVs, it is sufficient for finding an accurate wavelength solution that can then be input to \texttt{Telfit}. We choose to build an input wavelength solution this way as opposed to utilizing \texttt{Telfit}'s internal wavelength solution or the wavelength solution output by the IGRINS reduction pipeline, because both of the latter are built on line information from the HITRAN database\footnote{The IGRINS data reduction pipeline primarily calibrates its wavelength scales based on the OH night sky emission lines, but in regions with no significant OH lines, such as parts of the K band, the reduction pipeline also uses line information from the HITRANS database.} \citep{gord17}. HITRANS is usually accurate to better than the \ms level, but can be discrepant by 100 \ms or more \citep{roth05}, large enough to be unsuitable for this application.

Next, \texttt{Telfit} fits the A0 observation. By default, we only allow \texttt{Telfit} to vary parameters related to the abundances of the relevant molecular absorbers. For the orders used by \texttt{IGRINS RV}, these consist solely of CO, $\rm CH_{4}$, and $\rm H_{2}O$ in the K band, and $\rm CO_{2}$, $\rm CH_{4}$, and $\rm H_{2}O$ in the H band. Information on environmental conditions at the time of observation, such as the temperature and pressure, are also fed into \texttt{Telfit} and held fixed. Only if unknown are these allowed to vary.

\texttt{IGRINS RV} then takes the atmospheric parameters of the best fit and uses \texttt{Telfit} to generate a synthetic template comparable in resolution to the Livingston atlas. The \texttt{Telfit} fitting process also outputs the blaze fit, a seventh order polynomial, which is saved separately.

In generating this synthetic spectrum, \texttt{Telfit} is referencing its (potentially inaccurate) internal wavelength solution. We therefore calibrate this spectrum’s wavelength scale by comparing the wavelength scale of the Livingston atlas with \texttt{Telfit}'s best fit to the Livingston atlas itself. In other words, \texttt{IGRINS RV} applies \texttt{Telfit} to the Livingston atlas as it did to the A0 observation (and over the same wavelength range), takes the atmospheric parameters of the best fit and generates a synthetic version of the atlas, and then parameterizes any difference in the output wavelength scale. It does this by using \texttt{NLOpt} to fit the synthetic version of the Livingston atlas to the actual atlas, with only the wavelength solution allowed to vary. This fitted wavelength scale is the final wavelength solution associated with the synthetic telluric template generated. The accuracy of the synthetic telluric template's wavelength solution is thus directly anchored to that of the Livingston atlas.


For B frames, this entire process occurs exactly as just described, but for the A frames, the synthetic telluric template is constructed slightly differently. This is because the A frame spectra exhibit a slight dip in their blaze (Section~\ref{ssec:reduction}). As Telfit is only capable of modeling the blaze of a spectrum with a polynomial, its failure to account for this dip could lead to misfit absorption lines. \texttt{IGRINS RV} avoids this by running a spectral fit of the A frame telluric spectra (using its own spectral model) prior to applying Telfit. Because the blaze dip is properly paramterized by our spectral model, the pipeline can then use the best fit parameters for the dip to correct it out of the telluric data before they are input to Telfit\footnote{This dip correction itself could be inaccurate if the Livingston atlas were used as the telluric template, because the atlas is inflexible. \texttt{IGRINS RV} gets around this by generating the synthetic telluric templates for all of the B frames first (before the A frames) so that the B frame synthetic templates can be used in the fitting of the A frames that characterizes the blaze dip.}.
 
We expect the wavelength precision anchored to our telluric templates to be no greater than 10 to 20 \msnospace, as this is what past studies have found to be the typical RV stability of telluric lines \citep{seif08,fig10}.


\subsection{The Model Spectrum} \label{ssec:model}

Our spectral model intakes a pixel scale, a blaze correction, and 18 parameters (Table~\ref{tab:pars}). It begins by scaling the stellar and telluric templates through exponentiation (Table~\ref{tab:pars}, No. 1--2) and shifting the stellar template in wavelength space by its RV offset with respect to the telluric spectrum. It then reproduces the effects of rotational broadening on the stellar template with a convolution routine and re-bins both templates onto a common wavelength scale. 

The templates are combined and convolved with the instrumental profile (IP), whose provenance is discussed in Section~\ref{ssec:ip}. The broadened spectrum is then binned to a cubic wavelength scale and multiplied by a model for the blaze function of the observed data. The baseline for this blaze model is a seventh order polynomial fit to the associated telluric standard spectrum, sigma-clipped to avoid the skewing effects of absorption lines. This baseline is then combined with an optimizable polynomial (Table~\ref{tab:pars}, No. 12--18) to account for any slight differences between the telluric standard spectrum and the target data. The set degree of this polynomial varies depending on the size of the wavelength region being fit; in some cases it is as low as two degrees and in others as high as six. 

\begin{figure}[tb!]
\centering
\includegraphics[angle=0, width=1.\columnwidth]{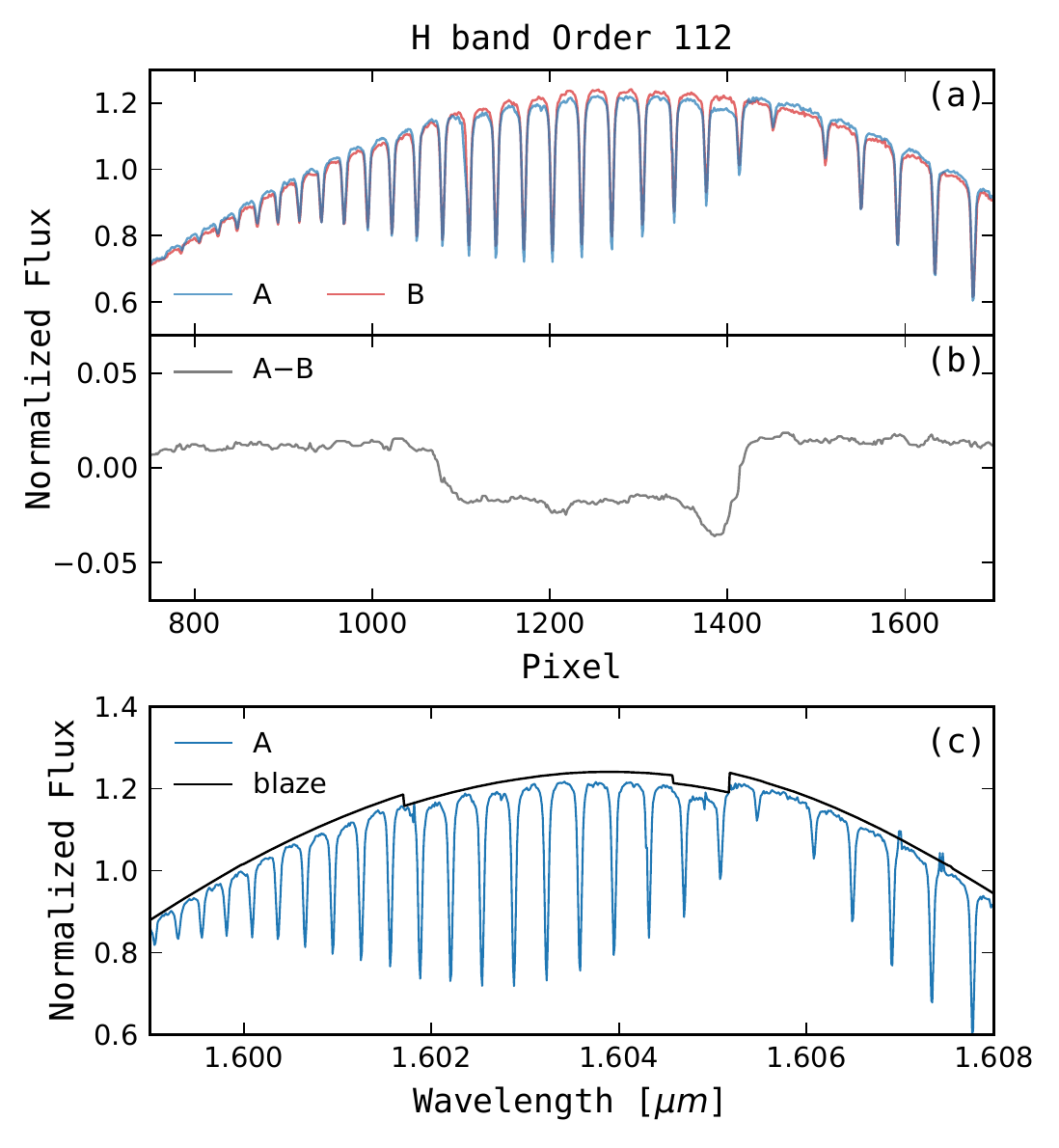}
\caption{
         (a) Comparison of A and B frame spectra of an A0 standard star, zoomed in near the center of order 112 in the H band.
         (b) The subtracted difference between the A and B frames. The characteristic shape of the effect is evident as a rectangular dip with a smaller, secondary dip descending from the rightmost edge. Overall, the effect is proportional (increasing with detector counts), discrete, present in both bands, and also occurs in data taken from Gemini South. 
         (c) The \texttt{IGRINS RV} best fit blaze model to the A frame spectrum, utilizing both a polynomial and the five-parameter dip model.
	    }
\label{fig:blazedip_fix}
\end{figure}

Lastly, if the spectrum is an A frame observation, a blaze correction is applied to the polynomial to account for the slight reduction in flux at the peak of the blaze (Figure \ref{fig:blazedip_fix}~c). This correction consists of a rectangular dip with a smaller, secondary rectangle descending from its rightmost edge. It is parameterized by five variables: the central location, width, and depth of the primary dip, and the width and depth of the secondary dip (Table~\ref{tab:pars}, No. 19--23). These are allowed to vary to some extent as part of the spectral fit. Were this feature not accounted for, \texttt{IGRINS RV} would measure a systematic difference between A and B frame RVs.

\begin{deluxetable}{c l}
\tablecaption{Model Parameters \label{tab:pars}
             }
\tabletypesize{\scriptsize}
\tablehead{ 
	 \colhead{No.}   & \colhead{Description}
	 }
\startdata
1--2 & Template scale factors\\
3 & Stellar template RV\\
4 & Stellar \vsini\\
5--7 & Quadratic Instrumental FWHM\\
12--14 & Quadratic blaze correction\\
15--18 & Addtl. blaze polynomial terms, as needed \\
19--23 & Flat field blaze correction (A frames only)
\enddata
\end{deluxetable} 


To fit the model spectrum to the data, we used the \texttt{NLOpt} \texttt{python} package \citep{john08} implementation of the bound Nelder-Mead optimization algorithm \citep{neld65,box65}. The optimizer starts from a set of initial parameter guesses and then moves through multiple cycles of steps, each separately fitting the blaze correction, the wavelength solution and telluric template power, the stellar template power and RV, the zero-th order IP width, and the \vsini (when applicable). At each step, the optimizer traverses the variable parameter space and calculates the chi-squared statistic in search of a minimum. To reiterate: the wavelength solution is ultimately calibrated by synthetic telluric templates generated from A0 observations, but the solution itself is always fit for on a spectrum by spectrum basis for every target.  The zero-th order IP width is fit similarly, with the higher orders fixed at values determined from a large scale analysis of A0 observations at different observatories and seasons (Section~\ref{ssec:ip}).

Barycentric velocity corrections are calculated using the \texttt{astropy} package \citep{exoplanet:astropy18} with coordinates and proper motions taken from {\it Gaia}\,DR2 \citep{gaia18}.

\subsection{Selection of Analysis Regions} \label{ssec:regions}

Although NIR RV monitoring is best suited for a late-type star like HD 189733 (a K2 dwarf), \texttt{IGRINS RV} can also be effectively applied to $\tau$\,Boo\,A (an F7 star). However, the high temperature of $\tau$\,Boo\,A means it exhibits little stellar content in the K band, so while both H and K bands were analyzed in the case of HD\,189733, only H band RVs were calculated for $\tau$\,Boo\,A.

Although IGRINS covers a wavelength range 1.05 $\mu$m wide, only those echelle orders with adequate stellar and telluric absorption lines are usable by \texttt{IGRINS RV}. Strong stellar lines are necessary for precise RVs and frequently spaced telluric lines are needed for an accurate wavelength solution. Through visual inspections of target star spectra, we found 10 orders in the H band (m = 100, 101, 102, 104, 111, 112, 114, 118, 119, 120) and 10 orders in the K band (m = 73, 74, 75, 76, 77, 78, 79, 81, 85, 87) which each exhibit adequate telluric absorption. From these, we found 5 orders in the H band (m = 104, 112, 114, 119, 120) were well-suited for Sun-like stars and 6 orders in the K band (m = 74, 75, 76, 77, 79, 81) were well-suited for late-type stars, in terms of stellar signal. Plots of these orders are provided in Appendix Figures~\ref{fig:GJ281_1}--\ref{fig:HD26257_2}.


We selected our K band wavelength regions by examining fits to both GJ\,281 (M0) and HD 189733 (K2) spectra. By using two stars of different spectral type, \vsini, and RV, observed at different airmasses, we ensured our selection is robust for a range of target star and atmospheric properties. We did the same with the H band, using HD 26257 (G2) and $\tau$\,Boo\,A (F7) for the diagnostic fits. In total, the stars studied range from F to M dwarfs.




\begin{deluxetable}{l l L}
\tablewidth{10em}
\tablecaption{Selected Wavelength Regions \label{tab:waves}
             }
\tabletypesize{\scriptsize}
\tablehead{ 
	 \colhead{Order (m)} & \colhead{$\rm Index^{*}$}  & \colhead{Wavelengths ($\rm \mu m$)}
	 }
\startdata
\multicolumn{3}{c}{H band}\\
\hline
104 & 6 & 1.7206-1.7295\\
112 & 14 & 1.6006-1.6075\\
114 & 16 & 1.5690-1.5831\\
119 & 21 & 1.5042-1.5061, 1.5122-1.5180^{\dagger}\\
120 & 22 & 1.4947-1.5065\\
\hline
\multicolumn{3}{c}{K band}\\
\hline
74 & 3 & 2.3880-2.3970\\
75 & 4 & 2.3565-2.3697\\
76 & 5 & 2.3260-2.3375, 2.3418-2.3472^{\dagger}\\
77 & 6 & 2.2965-2.3157, 2.3175-2.3224^{\dagger}
\enddata
\tablecomments{\\
    $^{*}$ The corresponding fits table layer index in IGRINS data, for user reference.\\
    $^{\dagger}$ Orders with multiple wavelength ranges selected will fit both simultaneously, masking the intervening wavelengths to ensure that they are not included in the $\chi^{2}$ goodness-of-fit calculation.
    }
\end{deluxetable} 


Beyond the requirement for well-spaced telluric and stellar absorption lines, we rejected 150 pixels on both edges of the detector to avoid poor fits resulting from the steep blaze and low S/N. We chose regions that are bookended by telluric absorption when possible, as this helps ground the cubic wavelength solution. Lastly, our visual inspections ensured no regions are included where the stellar template is a poor fit to the data, for example when an absorption line is present in an observation that is not in the line list of the stellar model. This last stipulation led us to reject orders 79 and 81 from the K band, leaving us with the final region selections shown in Table~\ref{tab:waves}. For some regions, we also masked parts of the middle of the order that contain a dearth of stellar information (e.g., 1.506--1.510 $\mu$m in order 119 of the H band). In total, we used $\sim$494~$\rm\AA$ of spectrum in the H band and $\sim$532~$\rm\AA$ in the K band.

\texttt{IGRINS RV} is provided with these regions predefined, but users can also customize their own lists of wavelength regions they would like the code to fit. This may be necessary if the user is studying target stars that are outside the range of spectral types presented here. As part of Step\,1, \texttt{IGRINS RV} will automatically take the input list of wavelength ranges and convert it into the echelle orders and pixel ranges that will be fit as part of RV estimates.


\begin{figure*}[t!]
\centering
\includegraphics[angle=0, width=1.\textwidth]{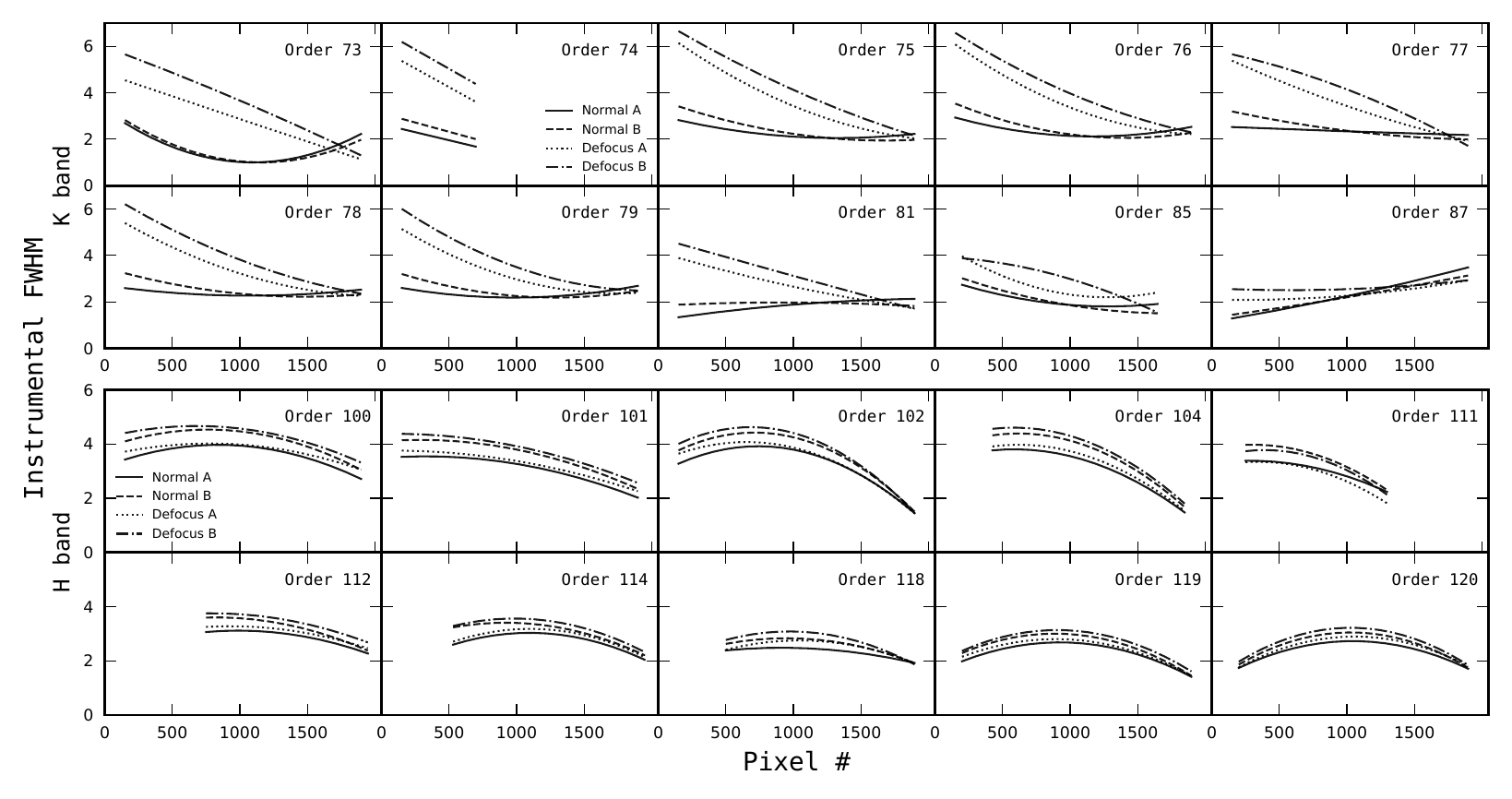}
\caption{
        The fitted solutions for instrumental profile width versus position on the detector in pixels, plotted for each echelle order with significant telluric absorption. The quadratic trend in resolution across the detector is evident, as is the the diverging behavior in the K band for observations taken during the defocus. The trends for orders 85 and 111 are inconsistent with the others as the result of misfitting; we discard these orders from our wavelength region selection process (they  feature a paucity of stellar lines regardless).
	    }
\label{fig:ip}
\end{figure*}


\subsection{Instrumental Broadening} \label{ssec:ip}

An accurate treatment of how resolution changes over the detector is essential for achieving the most precise RVs possible. While we found that a single Gaussian suitably models the IP at a given location within an echelle order (and the introduction of additional Gaussians leads to overfitting), the IP width can vary significantly across the detector. We employ a quadratic function in pixel space to parameterize these changes in each order. Instead of convolving narrower or wider Gaussians at each pixel location, we maintain the code's efficiency by stretching and compressing the pixel velocity scale such that convolution with a single, constant-width Gaussian produces the same results. The original pixel scale is still used for the remaining steps of producing the model spectrum.

We map the IP width variations through telluric standard fitting, which occurs as described in Section~\ref{ssec:tell}, with the important exception that now the linear and quadratic terms of the IP are allowed to vary. Each A0 star thus provides best-fit quadratics describing the IP variations across each order. We fit a total of 97 and 111 telluric standard observations in the H and K bands, respectively, with data taken over 5 years at McD at the DCT. 

Figure~\ref{fig:ip} shows the median IP fit from the ensemble of telluric standard fits, order by order, for the 10 orders with adequate telluric absorption in each of the H and K bands. Although the ensemble of fits are not pictured here for the sake of clarity, the IP curves from different observations appear largely consistent, supporting the choice of quadratics to capture the behavior (increasing the degree of the polynomial leads only to overfitting). 

The clear exceptions to this are the observations taken during the year of the K band defocus. As shown in Figure~\ref{fig:ip}, during this time most K band orders exhibit markedly broader IP widths, corresponding to lower spectroscopic resolutions, on their shorter-wavelength parts. In K band order 75, for example, the resolving power at pixel 200 decreases from roughly 53,500 to 25,000. At the higher-wavelength parts of the orders, on the other hand, the resolving power actually increases. These trends are in agreement with the behavior measured through the FWHM of sky OH emission lines (Figure~\ref{fig:defocus}).

We therefore undertook a separate treatment of the IP variations during this ``defocus" epoch. We also modeled the IP variations for A frame observations separately from B frame observations, as the two modes of observations orient the target star differently. We found that B frame observations exhibit similar IP variations to A frames except shifted to slightly higher IP widths; this minor difference is not unexpected given the frames involve different position of the target on the slit. 

For each frame type, A and B, and for each time period, the ``defocus" and the normal (i.e., all other times), the coefficients of these quadratics are hard-coded for use in all future fits. While the linear and quadratic terms are always held fixed at their hard-coded values, in order to take into account the possibility that different relative orientations between the star and the slit (among other things) may affect instrumental broadening, we allow the zeroth order IP width to vary slightly during the fitting of both telluric standards and target stars.

The diverging treatment of the ``defocus" time period here requires that all subsequent analysis consider observations from this period as statistically separate from those taken at other times. This has implications for estimating the pipeline’s overall precision and the uncertainty of individual RV data points.


\subsection{Calculating Final RVs and Uncertainties} \label{ssec:err}

With the completion of all the spectral fits in Step\,3, \texttt{IGRINS RV} yields an RV value for each individual order of each observation analyzed. For each exposure, \texttt{IGRINS RV} calculates two statistics from the RVs of its constituent observations: the mean ($\rm RV_{ij}$) and the standard deviation of the mean ($\rm \sigma_{ij}$), where ``i'' indexes different exposures and ``j'' indexes different orders. $\rm \sigma_{ij}$ contains information about the internal consistency of the RVs on a given night, and is typically around $\sim$13 \ms for GJ\,281 and $\sim$33 \ms for HD\,26257. For a full treatment of the uncertainty, we must measure the RVs' external consistency across a longer timescale of multiple nights using RV standards.

For an RV standard star, we calculate the variance of RVs within a given order across all exposures:

\begin{equation}\label{eq:sigmaj}
\rm \sigma_{j}^{2} = std(RV_{ij})^{2}\ for\ each\ order\ j.
\end{equation}
        
\noindent where ``std" refers to the standard deviation. $\rm\sigma_{j}^{2}$ thus provides information on the external RV precision of each order. However, because the scatter of RVs within an exposure also contributes to the scatter of RVs between different exposures, we must subtract out the internal uncertainty in order to determine the uncertainty in our analysis method:

\begin{equation}\label{eq:method}
\rm \sigma_{j(method)}^{2} = \sigma_{j}^{2} - median((\sigma_{i,j})^{2})\ for\ each\ order\ j,
\end{equation}

\noindent where $\rm\sigma_{j}^{2}$ is the exposure-by-exposure RV variance within an order (Eq.~\ref{eq:sigmaj}), and $\rm median((\sigma_{i,j})^{2})$ is the median of the variances of the different exposures. The $\rm\sigma_{j(method)}^{2}$ characterizes the uncertainty in our RV measurements as the result of inadequacies of the synthetic model to ideally represent the data, the RV instability of the telluric lines being used as a wavelength calibrator, and the intrinsic RV instability of the spectral region analyzed, among other systemic factors. For GJ\,281, we found $\sigma_{j}$ to range from 38--55 \ms (during the defocus, it ranged between 36--151 \msnospace), and $\sigma_{j(method)}$ to typically be around 40 \msnospace (85 \ms during the defocus). For HD\,26257, $\sigma_{j}$ ranged between 78--241 \msnospace, and $\sigma_{j(method)}$ was typically around 95 \msnospace.

Note that $\rm\sigma_{j(method)}^{2}$ can only be calculated by analyzing an RV standard star, as they do not have intrinsic RV variations. An RV standard is required to estimate $\rm\sigma_{j(method)}^{2}$ for each relevant echelle order in the H and K bands, as well as separately for the ``defocus" epoch. These uncertainties are built into the pipeline, facilitating accurate uncertainty estimates for targets that are not RV standards. In other words, analysis of a non-RV standard would skip Equations~\ref{eq:sigmaj} and \ref{eq:method}, instead using the $\rm\sigma_{j(method)}^{2}$ estimated from a previous analysis of an RV standard.

Using \texttt{IGRINS RV} in any way that significantly deviated from the form presented here (for example, deriving stellar templates from an alternate source) requires a repeat of this RV standard analysis in order to supply an estimate of $\rm\sigma_{j(method)}^{2}$ that accurately reflects the precision of the different method.

With a well-characterized uncertainty in the method, we can then add in a measure of the internal uncertainty for each exposure: the standard deviation of the mean of each exposure's constituent observations, $\rm\sigma_{ij}^{2}$. This provides information on uncertainties resulting from poorer quality or more rotationally broadened data. The total RV uncertainty associated with each exposure and order is thus the quadratic sum of both uncertainties:

\begin{equation} \label{eq:sigmaON2}
\rm S_{ij}^{2} = \sigma_{j(method)}^{2} + \sigma_{ij}^{2} 
\end{equation}

For each exposure, we calculate weights from $\rm S_{ij}^{2}$, which are then used to combine the different order RVs into one final RV and uncertainty. For each observation x,

\begin{equation}
\rm w_{j} = \bigg(\frac{1}{S_{xj}^{2}}\bigg)\bigg/\bigg(\sum_{j} \frac{1}{S_{xj}^{2}}\bigg)
\end{equation}

\begin{equation}\label{eq:finalRV}
\rm RV_{x} = \sum_{j} (w_{j} \cdot (RV_{xj} - \overline{RV_{xj}}))
\end{equation}

\begin{equation}
\rm \sigma_{x} = \left(\sqrt{\sum_{j} \frac{1}{S_{xj}^{2}}}\right)^{-1}
\end{equation}
\vspace{3pt}


Each set of RVs from a given order has its mean subtracted from it before the weighted combination in Eq.~\ref{eq:finalRV} occurs. This is because the orders have systematic RV offsets from one another (Section~\ref{ssec:abs}). The weighted mean acts linearly between orders, so the mean subtraction makes no difference for observations which have an RV estimated for all possible orders. Only for those observations that do not have an RV measurement from each order\footnote{A rare case that occurs if the corresponding data were too low quality, or there were too few observations to estimate a meaningful standard deviation, or because \texttt{Telfit} encountered a critical internal error when processing the night's A0 spectrum in that order.} does mean subtraction change the final RV calculated.


\section{Performance} \label{sec:results}

The typical run time of \texttt{IGRINS RV} depends on the amount of memory and number of CPU cores available on the user's machine. Using an Intel Core i9-9980XE CPU with 18 cores/36 threads, a run of all 64 nights and 4 orders of GJ\,281 in the K band took $\sim$6.6 hours for Step 1 (Telluric Modelling) and $\sim$1.2 hours for Step 3 (Analysis). As there are more orders to process in the H band, its runtimes are about 1.5 times as long.

\begin{figure}[tb!]
\centering
\includegraphics[angle=0, width=1.\columnwidth]{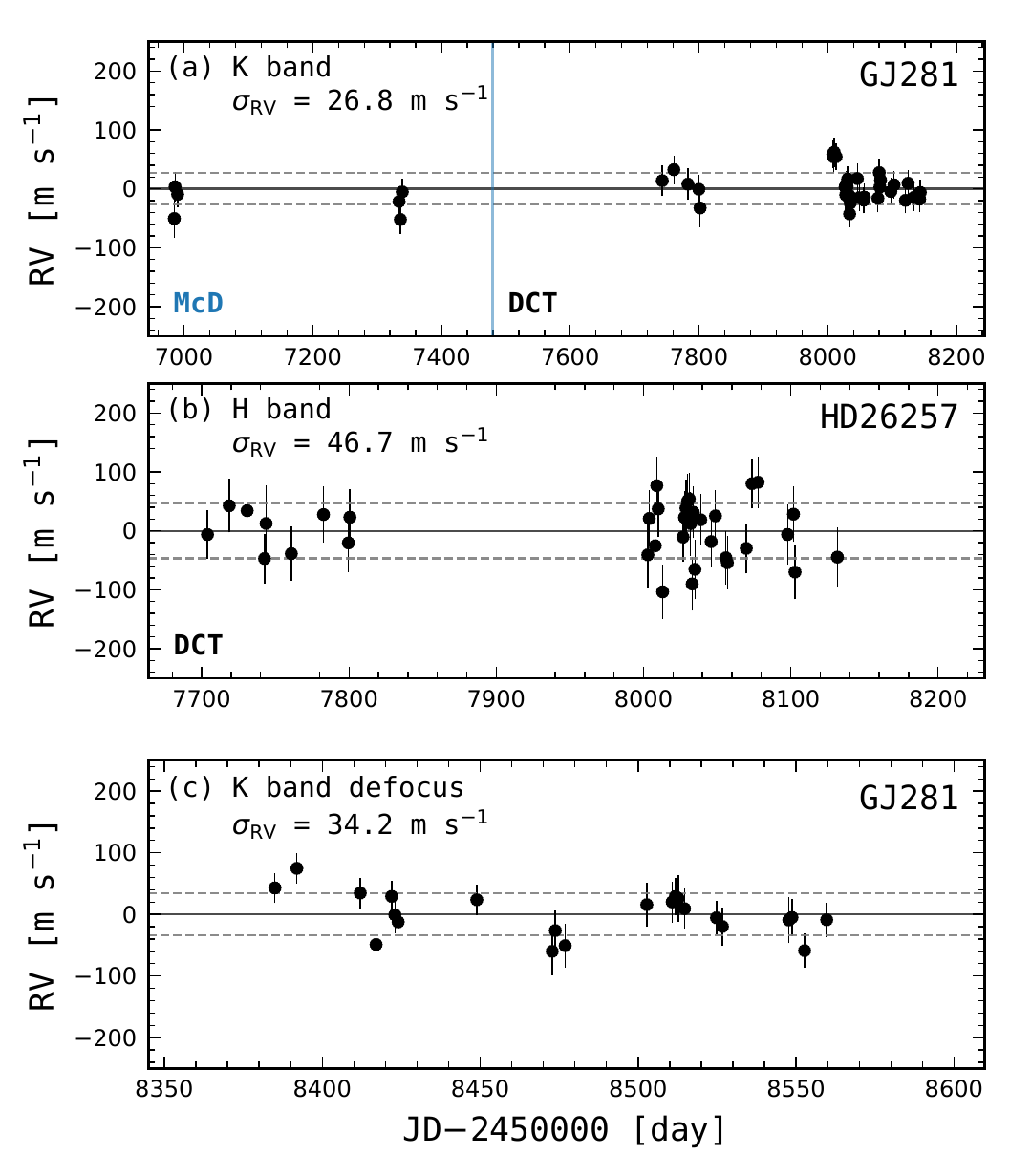}
\caption{
         Final RVs for the standard stars (a) GJ\,281 and (b) HD\,26257. The GJ\,281 observations span both McDonald and Lowell Observatory data, as well as data taken during the K band defocus. Dashed lines indicate one sigma deviations from the mean. 
	    }
\label{fig:RVstds}
\end{figure}


\begin{figure*}[p]
\centering
\includegraphics[angle=0, width=.95\textwidth]{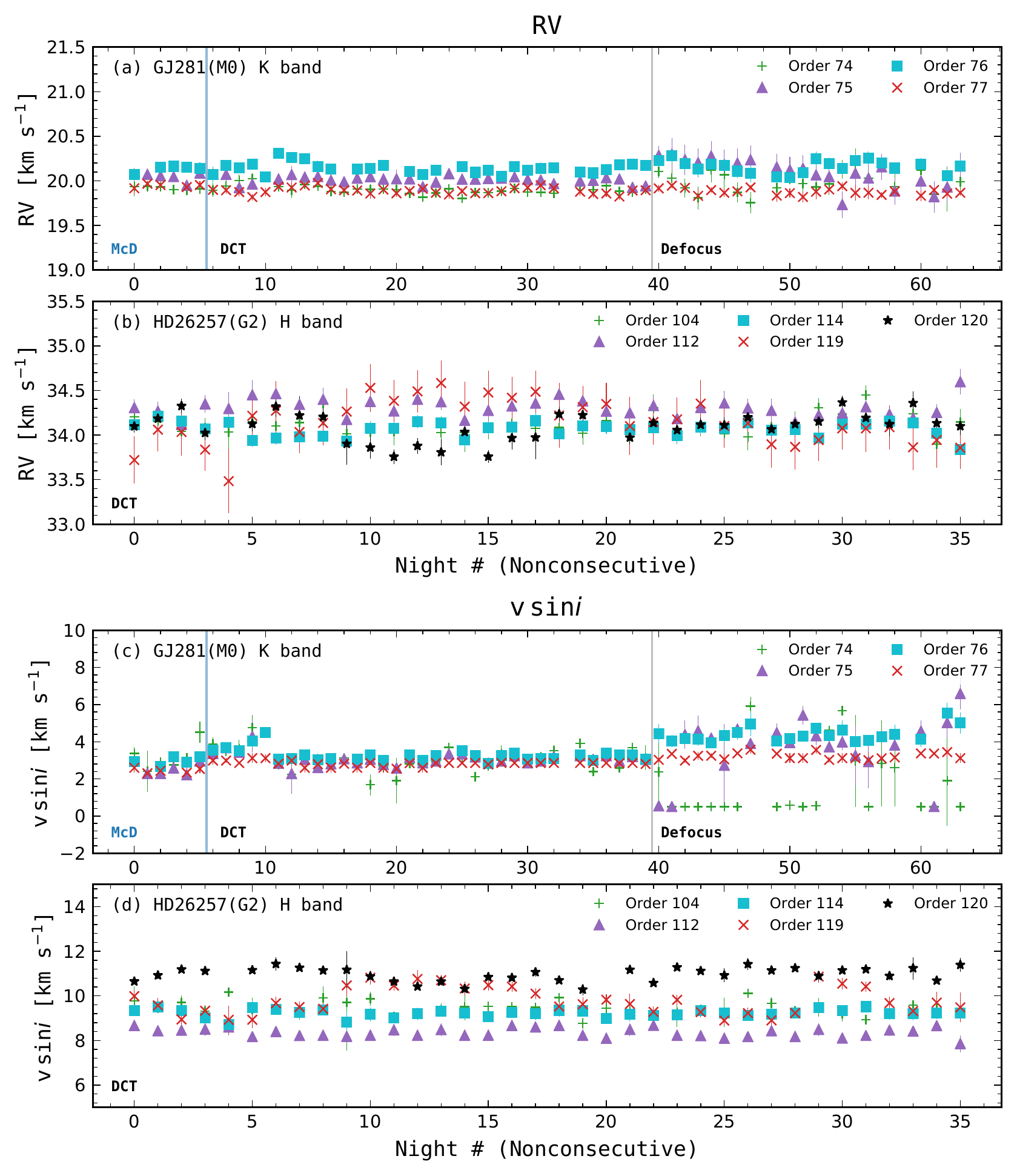}
\caption{
        Top panels: RVs for the standard stars GJ\,281 (a) and HD\,26257 (b) separated by order. Error bars are computed as the quadratic sum of the method uncertainty and the RV scatter within an exposure. Bottom panels: \vsini measurements separated by order for GJ\,281 (c) and HD\,26257 (d). Vertical axes have equal scaling for ease of comparison.
	    }
\label{fig:ObyO}
\end{figure*}

\subsection{RV Standards}\label{sec:results_STD}

We analyzed two different RV standards in order to estimate the precision delivered by \texttt{IGRINS RV}. Final output RVs for these targets are presented numerically in Table~\ref{tab:rv_result}. 

The results for GJ\,281 are shown in Figure~\ref{fig:RVstds} (a) and (c). The total variation over the yearlong K band defocus is 34.2 \msnospace, but over the three years of observations when the detector mounting was normally attached, the variation is 26.8 \msnospace. For the H band analysis of HD 26257, no such distinction between epochs is necessary because only the K band was defocused, and we estimate the single precision of 46.7 \ms across two years (Figure~\ref{fig:RVstds}~b). We consider the 26.8 \ms from GJ\,281 to be representative of the pipeline's precision in the K band; in the next section, we demonstrate better precision than 46.7 \ms is achievable in the H band for narrow-line (low \vsini) stars (Section~\ref{ssec:phosts}).

The uncertainty in each target's individual RVs agrees well with the overall RV variation measured. The average uncertainty of the GJ\,281 RVs is 26.5 \ms and the average uncertainty of HD\,26257's RVs is 46.8 \msnospace.

\texttt{IGRINS RV} also estimates the \vsini of the target star from its spectral model fits and converts these values into a single, final \vsini measurement. The procedure is almost exactly the same as used for determining an average final absolute RV (Section~\ref{ssec:abs}).

We estimate a \vsini of 2.96 $\pm$ 0.31 \kms for GJ\,281. This appears to be in rough agreement with the literature. \citet{scho19} measured the \vsini to be 2 \kms while \citet{hojj19} placed an upper limit on the \vsini of 2 \kmsnospace, but these values cannot be strictly compared with our own because they do not have uncertainties associated with them. Likewise, though our \vsini measurement of 9.24 $\pm$ 0.15 \kms for HD\,26257 may be somewhat overestimated compared to the literature values of 7.2 \kms \citep{gleb05} and 5.3 \kms \citep{bre16}, neither of the latter are reported with uncertainties.

There is some reason to expect \texttt{IGRINS RV} to have trouble determining \vsini in certain cases. The instrumental profile that IGRINS convolves with the stellar rotational profile is $\sim$\,7 \kms wide \citep{mace16}; for stars with rotational broadening much narrower than this, it may be difficult to measure \vsini accurately.

Figures~\ref{fig:ObyO} c and d show the \vsini results for GJ\,281 and HD\,26257 for a run of Step\,3 where \vsini is allowed to vary. We observe GJ\,281's \vsini estimates became significantly less precise and less accurate during the K band defocus, when the instrumental profile was much wider in some regions. Order 74 is a particularly good example. The analysis region of this order covers the part of the detector in which the resolution change manifested as a  severe decay (Figure~\ref{fig:ip}). As a result the \texttt{IGRINS RV} pipeline was not able to detect any of the rotational broadening of GJ\,281, often measuring a \vsini of 0 $\rm km\,s^{-1}$. The code is configured to automatically discard \vsini estimates from order 74 for observations taken during the defocus epoch.

Final runs of target stars should always fix \vsini at a constant value, given that the physical property presumably remains constant over the time frame of the of observations. The inaccuracy of \texttt{IGRINS RV} in estimating the \vsini of target stars can thus be expected to introduce some level of imprecision in the RVs. To test this, we executed a run of Step\,3 for each of our target stars in which \vsini was allowed to vary, providing an estimate of its value and an uncertainty in that estimate. We then ran Step\,3 three times: once with \vsini held fixed at this value, and once on a subset of nights with \vsini held fixed at this value plus one sigma and minus one sigma. We find an average absolute difference in RV of $<$1 \msnospace. We do not find any sign of higher \vsini inputs corresponding to higher RV outputs, or vice versa.

We caution users to perform this exercise for all target stars with high ($>$ 10 \kms) \vsini and/or those for which they require the most strictly accurate RV uncertainty estimates. Because the effect of \vsini on RVs appears uncorrelated, only one additional run with \vsini fixed at either one sigma greater or one sigma less than its estimated value will appropriately gauge the additional uncertainty term introduced as the standard deviation of the difference in RVs between the two runs. This additional uncertainty term is typically around 2--5 \msnospace, except for in the case of $\tau$\,Boo\,A, where it can be as high as $\sim$\,9 \ms (to be expected, given the higher \vsini of $\tau$\,Boo\,A). The new uncertainty can then be added in quadrature with the preexisting RV uncertainties, and an additional code module, Step 4, is provided as part of \texttt{IGRINS RV} to automate this calculation.

All RVs presented in this paper have had their uncertainties determined with this method.


\begin{deluxetable}{lRR c  lRR}
\tablecaption{Radial Velocity Estimates\label{tab:rv_result}
             }
\tablewidth{10pt} 
\tablehead{ 
	 \colhead{JD$-2450000$}  & \colhead{RV} & \colhead{RVerr} &&
	 \colhead{JD$-2450000$}  & \colhead{RV} & \colhead{RVerr} \\
	 \colhead{(day)} & \multicolumn{2}{c}{(m\,$s^{-1}$)}  &&
	 \colhead{(day)} & \multicolumn{2}{c}{(m\,$s^{-1}$)}  \\
     \cline{1-3} \cline{5-7}
     \colhead{(1)} & \colhead{(2)} & \colhead{(3)} &&
     \colhead{(1)} & \colhead{(2)} & \colhead{(3)} 
	 }
\startdata
\multicolumn{7}{c}{GJ\,281} 
\\\hline
6984.864046 &  -55.18 & 32.95 && 7339.012359 &   -9.93 & 23.07\\ 
6986.015578 &   -1.23 & 25.02 && 7742.946759 &    9.01 & 25.79\\ 
6990.013375 &  -14.12 & 22.13 && 7760.922781 &   27.70 & 24.24\\ 
7334.021467 &  -26.22 & 26.51 && 7782.803844 &    3.21 & 26.43\\ 
7336.043078 &  -56.77 & 23.93 && 7799.784772 &   -5.48 & 23.76
\\\hline
\multicolumn{7}{c}{HD\,26257} 
\\\hline
7703.895107 &   -7.14 & 41.93 && 7760.823794 &  -39.53 & 45.74\\ 
7718.691409 &   41.71 & 45.45 && 7782.671744 &   26.80 & 47.85\\ 
7730.849329 &   33.32 & 42.82 && 7799.594995 &  -21.31 & 49.66\\ 
7742.666440 &  -47.86 & 42.41 && 7800.586196 &   22.26 & 47.46\\ 
7743.701591 &   11.62 & 65.34 && 8002.918460 &  -41.56 & 55.13
\\\hline
\multicolumn{7}{c}{HD\,189733 H band} 
\\\hline
7145.832605 &   19.488 & 45.026 && 7296.710870 &  -70.167 & 46.943\\ 
7145.853588 &  -25.140 & 47.889 && 7296.715998 &  -62.021 & 47.380\\ 
7145.907048 &  -15.967 & 51.533 && 7296.721085 &  -67.346 & 47.992\\ 
7292.688153 & -242.442 & 44.407 && 7511.834885 &   16.084 & 42.277\\ 
7296.705488 &  -30.190 & 46.002 && 7511.848233 &  -16.432 & 42.554\\ 
\\\hline
\multicolumn{7}{c}{HD\,189733 K band} 
\\\hline
7145.832605 &   28.15 & 28.27 && 7296.710870 &    3.01 & 22.49\\ 
7145.853588 &  -34.69 & 25.76 && 7296.715998 &   -8.94 & 24.42\\ 
7145.907048 &  -67.53 & 33.72 && 7296.721085 &    1.27 & 22.67\\ 
7292.688153 & -158.41 & 22.99 && 7511.834885 &   21.75 & 21.87\\ 
7296.705488 &   33.18 & 25.27 && 7511.848233 &   19.95 & 21.99
\\\hline
\multicolumn{7}{c}{$\tau$\,Boo\,A} 
\\\hline
7029.929227 &  282.70 & 43.09 && 7049.965269 &   74.26 & 46.87\\ 
7029.934594 &  206.34 & 45.12 && 7049.969094 &   96.48 & 44.27\\ 
7029.991209 &  216.24 & 44.65 && 7049.973087 &  115.17 & 45.48\\ 
7049.957766 &  110.33 & 57.15 && 7049.976898 &   93.84 & 44.67\\ 
7049.961611 &  118.57 & 44.05 && 7050.030828 &   57.04 & 46.46
\enddata
\tablecomments{
    This table is available in its entirety in machine-readable form in the online journal.
    }
\end{deluxetable} 


\subsection{Validation of Planetary Systems} \label{ssec:phosts}

While monitoring RV standard stars allows a determination of \texttt{IGRINS RV}'s precision, only the recovery of an established stellar RV signal can test the code's reliability. We applied \texttt{IGRINS RV} to two stars with well-characterized planetary companions: HD\,189733 and $\tau$\,Boo\,A.



\begin{figure}[htb!]
\centering
\includegraphics[angle=0, width=1.\columnwidth]{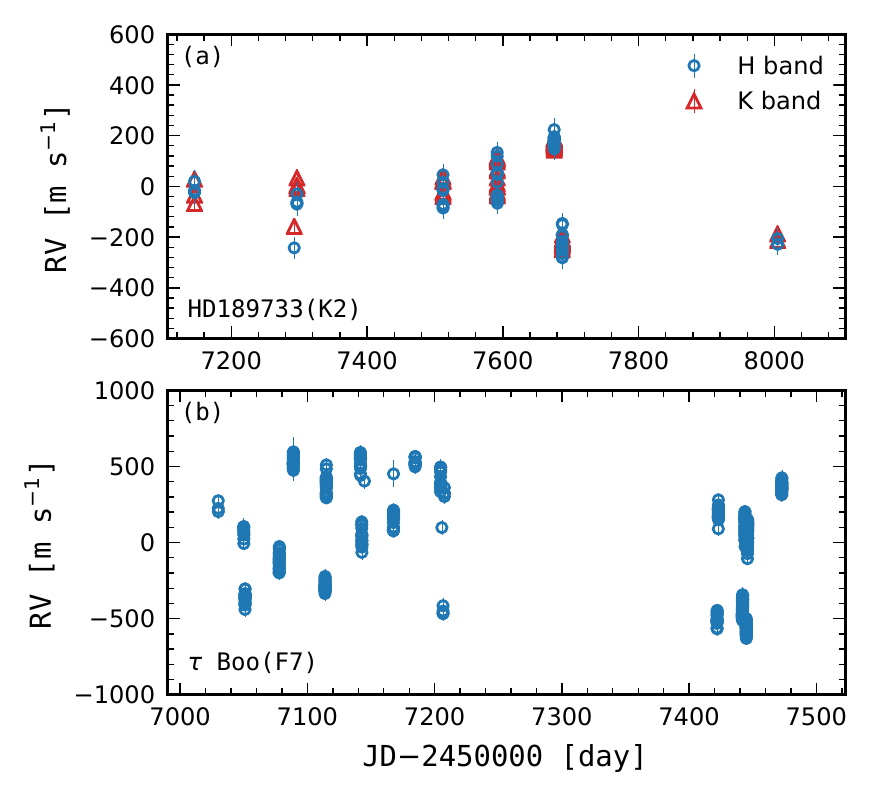}
\caption{
        RVs for (a) HD\,189733 and (b) $\tau$\,Boo\,A. 
	    }
\label{fig:RVJD}
\end{figure}

Output RVs are presented in Figure~\ref{fig:RVJD}, with numerical values available in Table~\ref{tab:rv_result}. Both stars exhibit RV dispersions much greater than the typical uncertainties in their measurements: 360.0 \ms vs. 46.9 \ms for $\tau$\,Boo\,A, 167.3 \ms vs. 43.1 \ms for HD\,189733 in the H band, and 158.4 \ms vs. 23.1 \ms for HD\,189733 in the K band. This indicated possible planet-induced reflex motion was occurring.

For each target, we searched for periodicity using the \texttt{astropy} \citep{exoplanet:astropy18} implementation of Lomb-Scargle periodograms, then obtained an orbital fit using \texttt{exoplanet} \citep{exoplanet:exoplanet} in combination with a maximum-likelihood optimization routine. The best fit parameters became the starting point of a Markov Chain Monte Carlo analysis, which we implemented with the \texttt{PyMC3} \citep{exoplanet:pymc3} tool\footnote{The orbital fit was done following the steps described on \url{https://docs.exoplanet.codes/en/stable/tutorials/intro-to-pymc3/}}. Four walkers were used, each with at least 20,000 steps and with the first 10,000 steps as burn-in.

\begin{figure}[htb!]
\centering
\includegraphics[angle=0, width=1.\columnwidth]{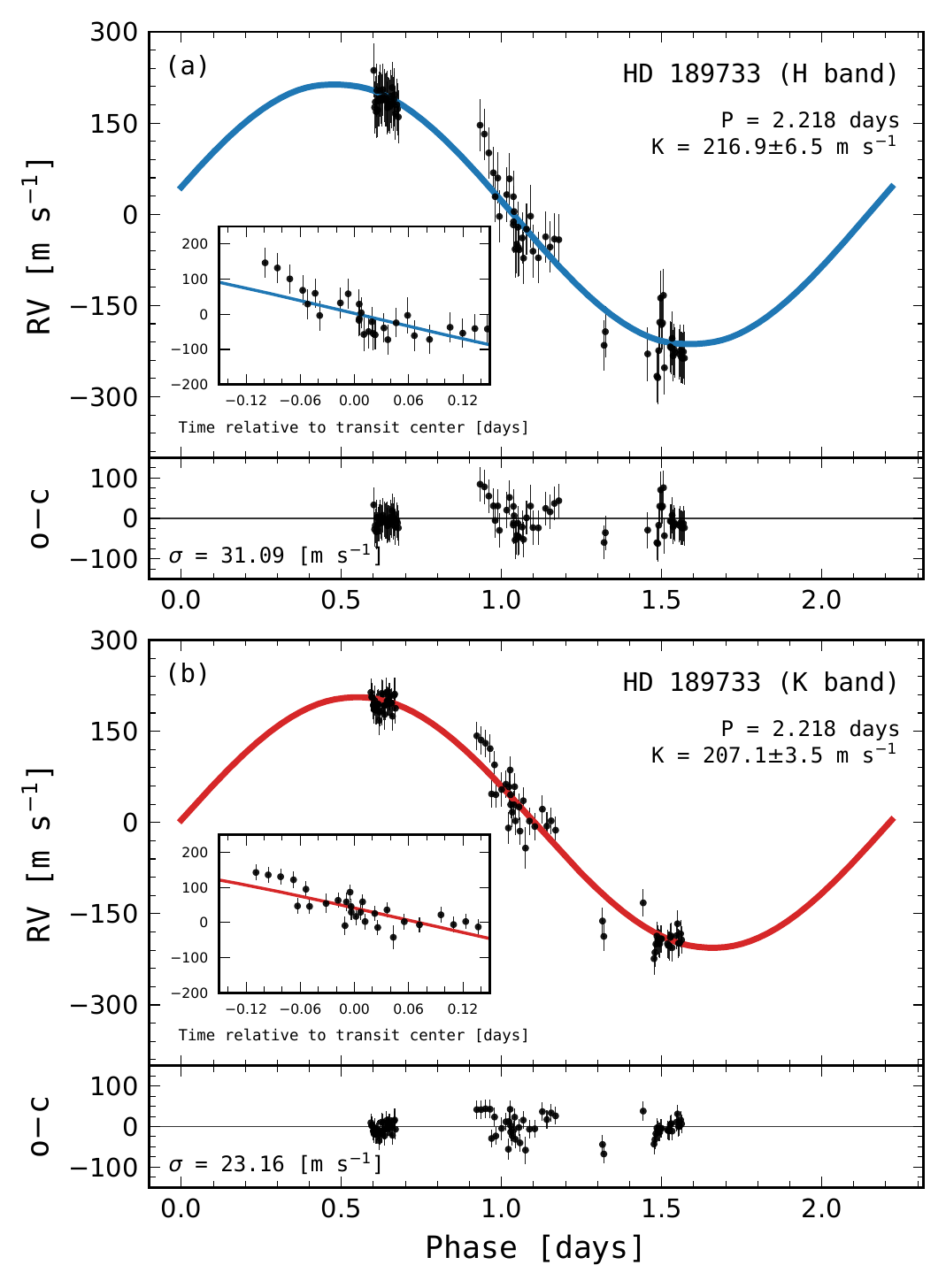}
\caption{
         HD\,189733 RVs for the H (a) and K (b) bands folded to the period of best fit. Transparent lines indicate the MCMC orbital fit; residuals from the mean fit are shown beneath. 
	    }
\label{fig:HD_orbitfit}
\end{figure}

\begin{figure}[htb!]
\centering
\includegraphics[angle=0, width=1.\columnwidth]{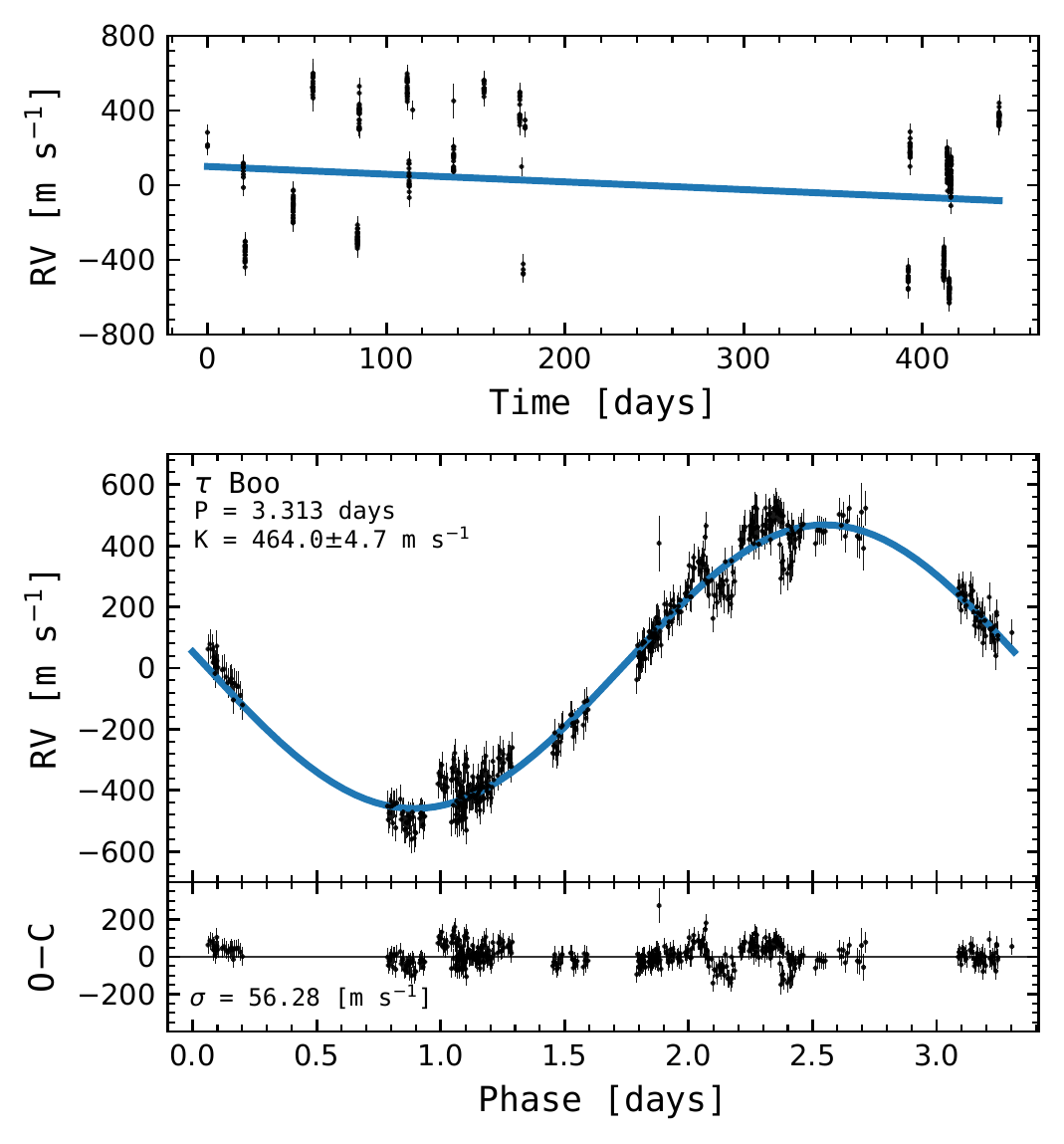}
\caption{Upper panel: $\tau$\,Boo\,A H band RVs alongside the best fit for the trend induced by the M dwarf companion. Lower panel: same RVs folded to the period of best fit, with line indicating the mean MCMC orbital fit, and residuals from the mean fit shown beneath.}
\label{fig:TauBoo_orbitfit}
\end{figure}

The planet-induced RV signals were recovered in both the H and K band data for HD\,189733, as well as in the H band data for $\tau$\,Boo\,A. Orbital fits are displayed in Figures~\ref{fig:HD_orbitfit} and \ref{fig:TauBoo_orbitfit}. All Keplerian fit parameters are in agreement with published values (Table~\ref{tab:fit_par}).

The fit to our measured RVs for $\tau$\,Boo\,A was slightly complicated by the presence of the companion $\tau$\,Boo\,B. Proceeding based off of Figure~3 of \citet{jus19} (the processed RVs were unavailable), we represented the M-dwarf companion's influence through a linear RV trend, which we fit for in combination with the planet-induced RV model. The result indicates that $\tau$\,Boo\,B produced a $-149.63\pm6.03$ \ms $year^{-1}$ slope (Figure~\ref{fig:TauBoo_orbitfit}, upper panel) in the RV of $\tau$\,Boo\,A over the course of our observations. This slope is comparable to the $\sim-100$ \ms $year^{-1}$ RV slope we estimate from \citet{jus19}.

Nearly one third of our RV measurements for HD\,189733 were taken during transit, and we indeed observe deviations in these RVs consistent with the Rossiter-McLaughlin effect (Figure~\ref{fig:HD_orbitfit}~a and b, panels). The semi-amplitude, $\sim50$ \msnospace, and the duration of the effect, $\sim$0.08 days, appear consistent with those of \citet{mout20} in their NIR observations of HD\,189733 with SPIRou (displayed in their Figure~4).

We did not analyze HD\,189733 with the intention of observing the Rossiter-McLaughlin effect, and consider its unexpected detection in our measured RVs to be further confirmation of the accuracy of the \texttt{IGRINS RV} code. Additionally, this indicates that the true scatter of the orbital fit residuals for HD\,189733 are likely smaller than measured, as \citet{mout20} found that including the Rossiter-McLaughlin effect in their model during MCMC fitting reduced the scatter of their residuals by an average of 34\%.

The scatter of the orbital fit residuals compares favorably with that of the RV standards, while also tracing the effects of \vsini on precision. In the K band, the residuals for HD\,189733, 23.2 \msnospace, exhibited similar scatter as we saw in the RV standard GJ\,281, 26.8 \msnospace. Both stars have a comparably small \vsini of 2--4 \kmsnospace. In the H band, the standard deviation of the HD\,189733 orbital fit residuals was 31.1 \msnospace, smaller than the precision obtained from the RV standard HD\,26257, 46.7 \msnospace. This is explained by the faster rotation of HD\,26257 (\vsini$\sim$10 \kms) compared to HD\,189733 (\vsini$\sim$4 \kms). The stronger rotational broadening produces shallower stellar absorption lines, which in turn is associated with lower RV information content \citep{butl96}. 

This trend is also seen in $\tau$\,Boo\,A, which has an even higher \vsini of $\sim$20 \kmsnospace, and delivers a standard deviation around its orbital fit of 56.28 \msnospace. This target was highly saturated in the IGRINS slit-viewing camera during observations, making it difficult to center the star in the slit and ensure it was in the same place for every observation. It is therefore possible that a changing slope in the slit illumination resulting from guiding and/or centering variations might have contributed to this higher error, not just the \vsini.

Taken as a whole, these results suggest that the true baseline precision \texttt{IGRINS RV} is capable of in the H band is actually 31.1 \msnospace, as opposed to 46.7 \msnospace, for stars with relatively low \vsini ($\lessapprox$ 5 \kmsnospace).

The fact that the precision changes with \vsini does not mean that \texttt{IGRINS RV} underestimates its output RVs uncertainties. If a spectrum contains less RV information content, this will be reflected in a higher scatter for the RVs calculated from different exposures within an observation, producing final RV uncertainties that are appropriately higher, as well. This behavior is discernible in the RVs of the planet hosts, where the uncertainties increase proportionally to the scatter around the true RV signal such that $>$1 sigma outliers appear less than 32\% of the time, while $>$2 sigma outliers occur less than 5\% of the time.

\subsection{Template Choice} \label{ssec:disc_templates}

\begin{figure*}[htb!]
\centering
\includegraphics[angle=0, width=1.\textwidth]{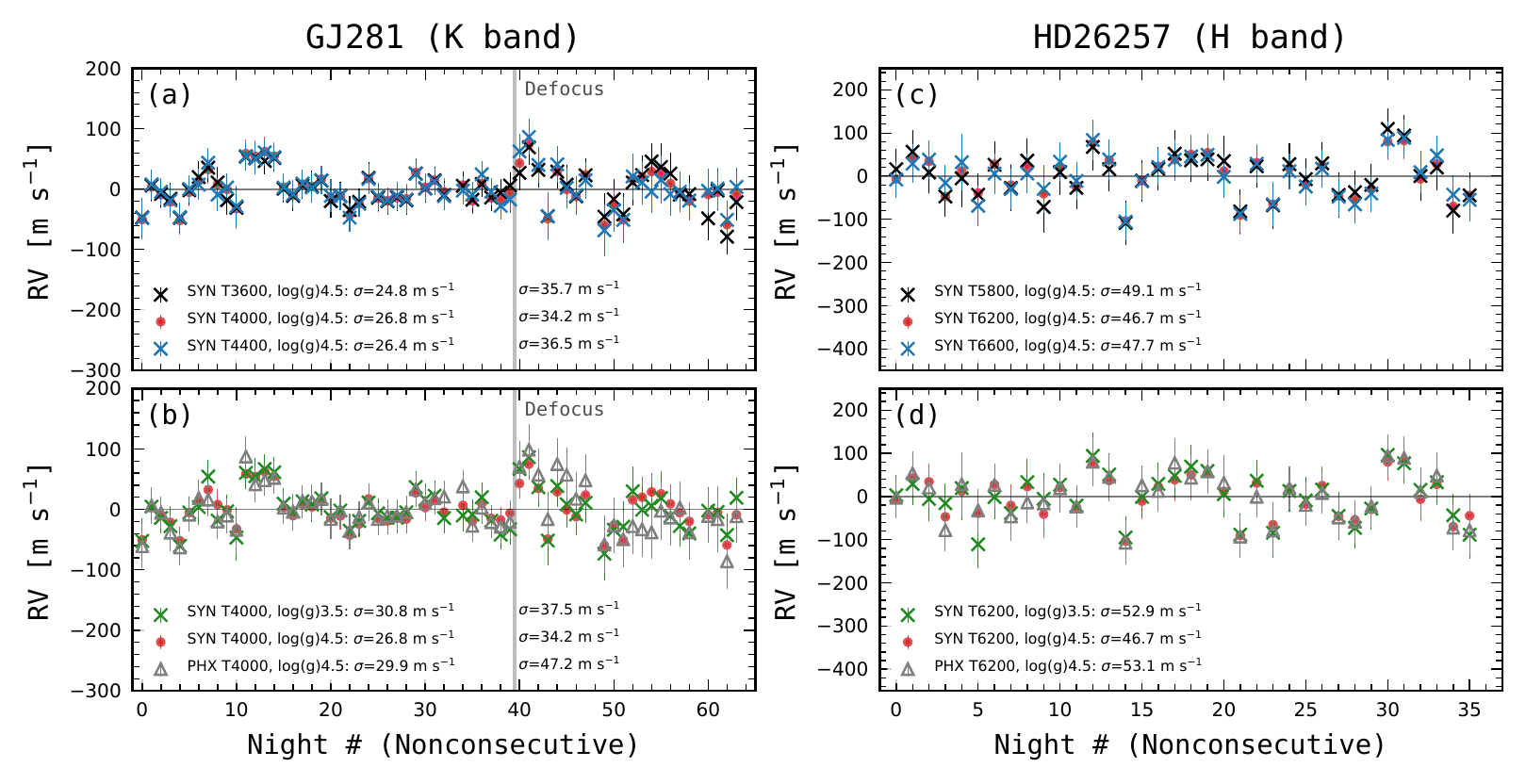}
\caption{
         RVs for the standard stars GJ\,281 and HD\,26257 using stellar templates with different effective temperatures (a) and (c) and \logg (b) and (d). Panels (b) and (d) also display results calculated with a template constructed with the same stellar parameters, but different software. SYN refers to our synthetic stellar templates produced with VALD and PHOENIX NextGen, and PHX refers to stellar templates from the G\"{o}ttingen Spectral Library \citep{huss13}. 
        }	  
\label{fig:Tlogg}
\end{figure*}

As shown in Figures~\ref{fig:Tlogg} (a) and (c), moderate ($\sim$ 400\,K) mismatches in T$_{\rm eff}$ between the chosen stellar template and the target star have little effect on the \texttt{IGRINS RV} results. For GJ\,281 and HD\,26257, such template changes amount to only a few \ms difference in precision at most. Mismatches in \logg appear to have similarly small effects (Figures~\ref{fig:Tlogg}~b and d). In addition, we tested \texttt{IGRINS RV} with stellar templates from the G\"{o}ttingen Spectral Library\footnote{\url{http://phoenix.astro.physik.uni-goettingen.de/?page_id=15}. These spectra require flattening before use.} \citep[Figure~\ref{fig:Tlogg}~b and d;][]{huss13}. 

Unlike with \vsini or average RV, \texttt{IGRINS RV} does not come with any built-in functionality to converge to more accurate stellar template parameters. The only way for the user to check their choice of stellar template is to run Step 2 or 3 with stellar templates of different parameters and compare their output uncertainties and model fit plots.

When selecting a stellar template, the more closely the stellar template resembles the target star, the more precise the results will be. To some extent, worse spectral fits resulting from a less accurate stellar template will be reflected in a higher scatter for the RVs calculated from different exposures within an observation, meaning the final RV uncertainty estimates will be appropriately higher as well.

However, if using stellar templates produced by a means other than the one presented here, there is no guarantee that the mismatch behavior will fully encapsulate the intra-exposure uncertainty introduced. Only another analysis of an RV standard would provide a robust quantification of the precision of the code in this context. For instance, Figures~\ref{fig:Tlogg} b and d show that running our RV standards with synthetic stellar templates produced with a different model atmosphere code and line list \citep{huss13} results in lower precision. In the K band, the increased scatter occurs in tandem with increased intra-exposure uncertainty (the error bars), but in the H band it does not, such that the full inter-night scatter only calculable from an RV standard is required for an accurate characterization of the precision. 



\begin{deluxetable}{l cc}
\tablecaption{Keplerian Fit Parameters\label{tab:fit_par}
             }
\tabletypesize{\scriptsize}
\tablehead{
	 \colhead{} & \colhead{This study}  & \colhead{Literature}
	 }
\startdata
\multicolumn{3}{c}{HD189733 H band} 
\\\hline
Period [days]                       &   2.218     & 2.219$\pm$0.0005  \\
Orbital eccentricity                &    0.0028 [fixed]$^{\dagger}$& 0 [fixed]         \\
RV semi-amplitude [\ms]             &  216.86$\pm$6.55      & 205$\pm$6  
\\\hline
\multicolumn{3}{c}{HD189733 K band} 
\\\hline
Period [days]                       &   2.218     & 2.219$\pm$0.0005  \\
Orbital eccentricity                &    0.0028 [fixed]$^{\dagger}$& 0 [fixed]         \\
RV semi-amplitude [\ms]             &  207.21$\pm$3.54      & 205$\pm$6  
\\\hline
\multicolumn{3}{c}{$\tau$\,Boo} 
\\\hline
Period [days]                       &   3.313     & 3.3124568$\pm$0.0000069	\\
Orbital eccentricity                &   0.014$\pm$0.008     & 0.011$\pm$0.006           \\
RV semi-amplitude [\ms]             &  464.02$\pm$4.68      & 471.73$\pm$2.97           \\
RV trend [\ms yr$^{-1}$]            &$-$150.82$\pm$6.07     &   \nodata
\enddata
\tablecomments{Literature values for HD189733 and $\tau$\,Boo are from \citet{bouc05} and \citet{bors15}, respectively.\\
$^{\dagger}$ From transit data \citep{ball19}. Uncertainties on periods from this study are negligibly small.}
\end{deluxetable} 

\subsection{Absolute RVs} \label{ssec:abs}

\texttt{IGRINS RV} can also deliver absolute -- as opposed to relative -- RVs. The absolute RVs measured from different orders are usually consistent with each other within uncertainties, as are the means of the RVs from each order (Figures~\ref{fig:ObyO}~a and b). However, on occasion one or more orders exhibit zero-point offsets in RVs from the rest (e.g. order 76 in the GJ\,281 analysis). These relative offsets between orders are consistent for a given target, but not between targets. The behavior persists to some extent in all observations, and in some cases the discrepancies are larger than those shown here.

A wide array of diagnostic tests was undertaken to determine whether these offsets were caused by an inadequate treatment of true variations between echelle orders in the spectra. Ultimately, we found that the effect is inconsistent with a systematic origin, such as a wavelength calibration error or an error in instrumental profile modeling, and is instead best explained by slight inaccuracies in the depths and/or wavelengths predicted for the absorption lines in either our stellar models, telluric models, or both. Future versions of \texttt{IGRINS RV} will explore means of reducing such discrepancies.

As the effect only influences the zero-point offsets of the RVs derived from each echelle order, it can be mitigated in a relatively straightforward manner. Although a more involved correction could be applied, such as the Trend Filtering Algorithm used in \citet{cale18} to detrend order-dependent RVs, we found that subtraction of the mean RV from each order before weighted combination (Eq.~\ref{eq:finalRV}) successfully mitigated the effect without introducing the risk of overfitting our results. Because the relative RV variations are precise, the pipeline's capabilities when it comes to RV monitoring experiments and planet detection are unaffected.

\texttt{IGRINS RV} still provides the ability to compute absolute RVs. In this setting, the discrepancy between the RV zero-points of different orders is treated as a source of uncertainty in itself, which is added into the final RV uncertainties. In some cases, this added uncertainty will be small, and the precision of absolute RV measurements will be comparable to those reported here for relative RVs, but this is not guaranteed. The absolute RV is therefore provided in support of science cases other than planet detection, in which \kms characterization of an absolute RV is all that is required
\citep[e.g., star cluster and moving group member identification and star cluster dynamic studies; ][]{tan19, pan20}. The absolute RVs \texttt{IGRINS RV} derived for our RV standards (Table~\ref{tab:info}) exhibited slightly less precision, with mean RV uncertainty increasing to 30--45 \ms for GJ\,281 and 53 \ms for HD\,26257. Both of these RVs were relatively accurate, as well: we found the absolute RV of GJ\,281 in both the normal and defocused epochs to be within uncertainty of the value provided by \citet{mald10}, $20.23\ \pm\ 0.26$ \kmsnospace. For HD\, 26257, although the pipeline's absolute RV disagrees with the value of $33.649\ \pm\ 0.0055$ \kms reported by \citet{soub18}, the discrepancy is only 0.267 \kmsnospace. This suggests \texttt{IGRINS RV} delivers absolute RVs good to a systematic uncertainty of a few hundred \msnospace.

\subsection{Suggested Observing Strategy}

To optimize RVs calculated from IGRINS observations, telluric standard observations are a requirement and should be taken immediately before or after science target observations (and within $\sim$~0.3 in airmass). Telluric standards observed at more discrepant airmasses can still be used, albeit with caution. If the characteristics of the atmosphere remained the same between the standard and science target observations, then the spectral model's ability to adjust the telluric template's power should adequately compensate for a relatively significant difference in airmass. However, if the atmosphere changed significantly over the course of the night (e.g. humidity increased), then the telluric template generated from the standard may poorly correspond to the absorption featured in the science target spectra. This would result in model misfits and, potentially, poor RVs. If users use telluric standards at discrepant airmasses, we suggest that they refer to the relevant weather logs and to check the fit plots outputted by \texttt{IGRINS RV} for the observation(s) in question.

As the atmosphere can be highly variable from night to night, we discourage trying to substitute telluric standard observations from different nights. Other than telluric standards, there is no need to take additional calibrations beyond what is required for typical IGRINS observing.

If the user is constructing stellar templates following a different procedure than that presented here, employing alternate wavelength regions for analysis, or otherwise altering the pipeline in ways that would systematically affect its performance, observations of an RV standard star are required to characterize the precision of the alternate method. We suggest RV standards be narrow-line stars observed repeatedly across multiple epochs.

Lastly, we recommend users observe in a nodding sequence of at least four frames (i.e., ABBA as opposed to AB). This allows for better characterization of the internal RV uncertainty within a given night of observation. If users are unsure what exposure time will yield their desired S/N, we recommend first taking one AB pair and measuring its S/N on the fly. The user can then decide whether they need to take only one more AB pair, or to adjust the exposure time and take two.





\section{Summary and Future Prospects} \label{sec:conc}

We present \texttt{IGRINS RV}, an open source \texttt{python} pipeline for computing RV measurements from reduced IGRINS spectra. Although IGRINS was not designed to produce precise RVs, by applying a modified forward modeling technique to the cross-dispersed echelle spectrograph format, we find that the wide wavelength range of IGRINS allows us to achieve high precision measurements. 

With this large wavelength range --- the full H and K bands --- comes a variety of challenges. Instrumental profile variations across the detector are significant and require extensive modeling, especially given a yearlong period when the K band was defocused. Furthermore, a large portion of the H and K bands have inadequate information content to deliver accurate RVs, such that the regions used by \texttt{IGRINS RV} had to be handpicked based on a variety of criteria. We ultimately found a total of $\sim$494 $\rm\AA$ in the H band suitable for the analysis of an F7 star, a G2 star, and a K2 star ($\tau$\,Boo\,A, HD\,26257, and HD\,189733, respectively); $\sim$532 $\rm\AA$ in the K band were used in the analysis of an M0 star and a K2 star (GJ\,281 and HD\,189733).

Perhaps most importantly, telluric absorption is extremely variable, and static templates are not viable over these large wavelength ranges. Instead, we use a combination of \texttt{Telfit} and our own forward-modeling code to construct accurate, high-resolution telluric templates on a night by night basis. This provides a common-path wavelength calibrator while also alleviating the need to mask or correct for telluric lines. 

\texttt{IGRINS RV} requires no additional instrument hardware, such as a gas cell or laser frequency comb. Only accompanying observations of telluric standard stars are a prerequisite. 

The \texttt{IGRINS RV} pipeline succeeds in achieving nearly 10 times better precision than the method previously applied to measure RVs with IGRINS \citep{mace16}, demonstrating a precision in the K band of 26.8 \ms when applied to the narrow-line star GJ\,281 (\vsini$\sim$3 \kmsnospace) and 31.1 \ms in the H band when applied to HD\,189733 (\vsini$\sim$4 \kmsnospace). These precisions are validated by a monitoring campaign of RV standard stars and the successful retrieval of the planet-induced RV signals of HD\,189733 and $\tau$\,Boo\,A from actual observational data.

\texttt{IGRINS RV} has been tested successfully with Gemini South data. Though its performance on Gemini South observations has not been characterized with an RV standard, we expect the capabilities reported here to extend to such data. The RVs (Figures~\ref{fig:ObyO}~a and b), stellar rotational velocities (Figures~\ref{fig:ObyO}~c and d), and instrumental profile patterns (Figure~\ref{fig:ip}) all appear consistent in accuracy and precision between observations taken at McD and DCT. Observations taken from Gemini South may actually produce better quality RVs than those presented here, as the smaller slit (0.34\arcsec) may reduce variations in slit illumination.

Future versions of \texttt{IGRINS RV} will explore means to further improve the accuracy of our stellar and telluric models in order to reduce the zero-point RV discrepancies measured between echelle orders. Additionally, a new version of \texttt{IGRINS RV} will be released once the IGRINS plp is updated to address the flux suppression effect observed in reduced A frames.

\texttt{IGRINS RV} comes with a high degree of built-in functionality and is designed for ease of use. The software can be run entirely from the command line, and should the user choose to modify it in any way, the code is extensively commented. For full instructions on how to run \texttt{IGRINS RV}, a detailed instruction manual is available on the wiki page at the GitHub repository \citep[][submitted]{tang21}. 

NIR RVs provide a valuable tool for planet detection around spotted and active stars, especially T Tauri stars, because the lower impact of stellar activity at longer wavelengths can distinguish between apparent RV variation arising from cool spots and planetary companions. At the same time, M-dwarfs, important targets for exoplanet searches \citep{fran16}, yield the best signal to noise in the NIR because they emit most of their light at longer wavelengths. The \texttt{IGRINS RV} pipeline facilitates this valuable application for an already powerful spectrograph. Moreover, our code provides a proof-of-concept for a precise RV methodology that can be developed for other NIR echelle spectrographs not necessarily designed for such experiments.


\acknowledgments
We are grateful to the anonymous referee for a timely, constructive, and insightful report that improved the quality and presentation of our manuscript. We thank the technical and logistical staff at McDonald and Lowell Observatories for their excellent support of the Immersion Grating Infrared Spectrograph (IGRINS) installations, softwares, and observation program described here. In particular, D. Doss, C. Gibson, J. Kuehne, K. Meyer, B. Hardesty, F. Cornelius, M. Sweaton, J. Gehring, S. Zoonematkermani, E. Dunham, S. Levine, H. Roe, W. DeGroff, G. Jacoby, T. Pugh, A. Hayslip, and H. Larson. We also thank Laura Flagg for her insightful conversations. Partial support for this work was provided by NASA Exoplanet Research Program grant 80-NSSC19K-0289 to L. Prato. CMJ would like to acknowledge partial support for this work provided through grants to Rice University provided by NASA (award 80-NSSC18K-0828) and the NSF (awards AST-2009197 and AST-1461918). We are grateful for the generous donations of John and Ginger Giovale, the BF Foundation, and others which made the IGRINS-DCT program possible. Additional funding for IGRINS at the DCT was provided by the Mt. Cuba Astronomical Foundation and the Orr Family Foundation. IGRINS was developed under a collaboration between the University of Texas at Austin and the Korea Astronomy and Space Science Institute (KASI) with the financial support of the US National Science Foundation under grant AST-1229522 and AST-1702267, of the University of Texas at Austin, and of the Korean GMT Project of KASI.

This work has made use of the VALD database, operated at Uppsala University, the Institute of Astronomy RAS in Moscow, and the University of Vienna. This study also made use of the SIMBAD database and the VizieR catalogue access tool, both operated at CDS, Strasbourg, France. These results made use of the Lowell Discovery Telescope (LDT) at Lowell Observatory. Lowell is a private, non-profit institution dedicated to astrophysical research and public appreciation of astronomy and operates the LDT in partnership with Boston University, the University of Maryland, the University of Toledo, Northern Arizona University and Yale University. We have used IGRINS archival data older than the 2 year proprietary period.


\vspace{5mm}
\facilities{Lowell Discovery Telescope (LDT, formerly the Discovery Channel Telescope, DCT), 
            McDonald Observatory (McD)}

\software{\texttt{astropy} \citep{exoplanet:astropy18},  
          \texttt{exoplanet} \citep{exoplanet:exoplanet},
          \texttt{pymc3} \citep{exoplanet:pymc3},
          \texttt{theano} \citep{exoplanet:theano},
          \texttt{The IvS Python Repository} \citep{ivs},
          \texttt{BMC} \citep{duar18},
          \texttt{Telfit} \citep{gull14},
          \texttt{NLOpt} \citep{john08}.
          }


\appendix{}
\counterwithin{figure}{section}
\counterwithin{table}{section}


\section{Spectra Plots for RV Standards}

\begin{figure*}[p]
\centering
\includegraphics[angle=90, width=0.8\textwidth]{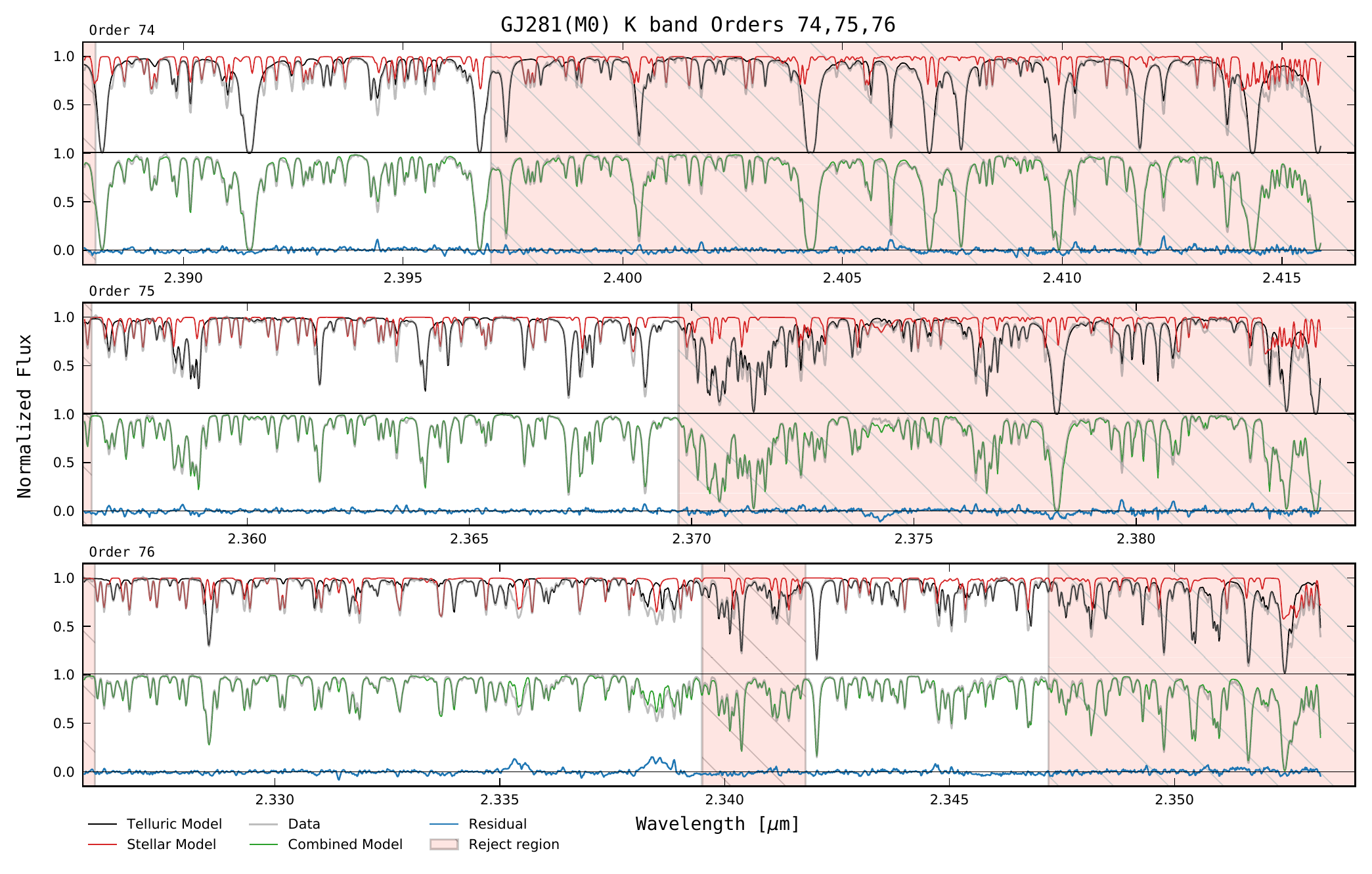}
\caption{
        Wavelength regions selected for K band RV analysis, with GJ\,281 as an example star. Orders 79 and 81 in the K band did not make it into the final selection, as they only exhibited small regions of significant, interspersed stellar and telluric absorption. Crossed-out regions were rejected or masked based off the criteria listed in Section \ref{ssec:regions}, with the exception of the large area of order 74 that was rejected because of improper telluric fitting. The regions displayed were trimmed by 150 pixels on each side to remove the worst of the blaze effects. Figure continues on next page.
	    }
\label{fig:GJ281_1}
\end{figure*}

\begin{figure*}[p]
\centering
\includegraphics[angle=90, width=0.8\textwidth]{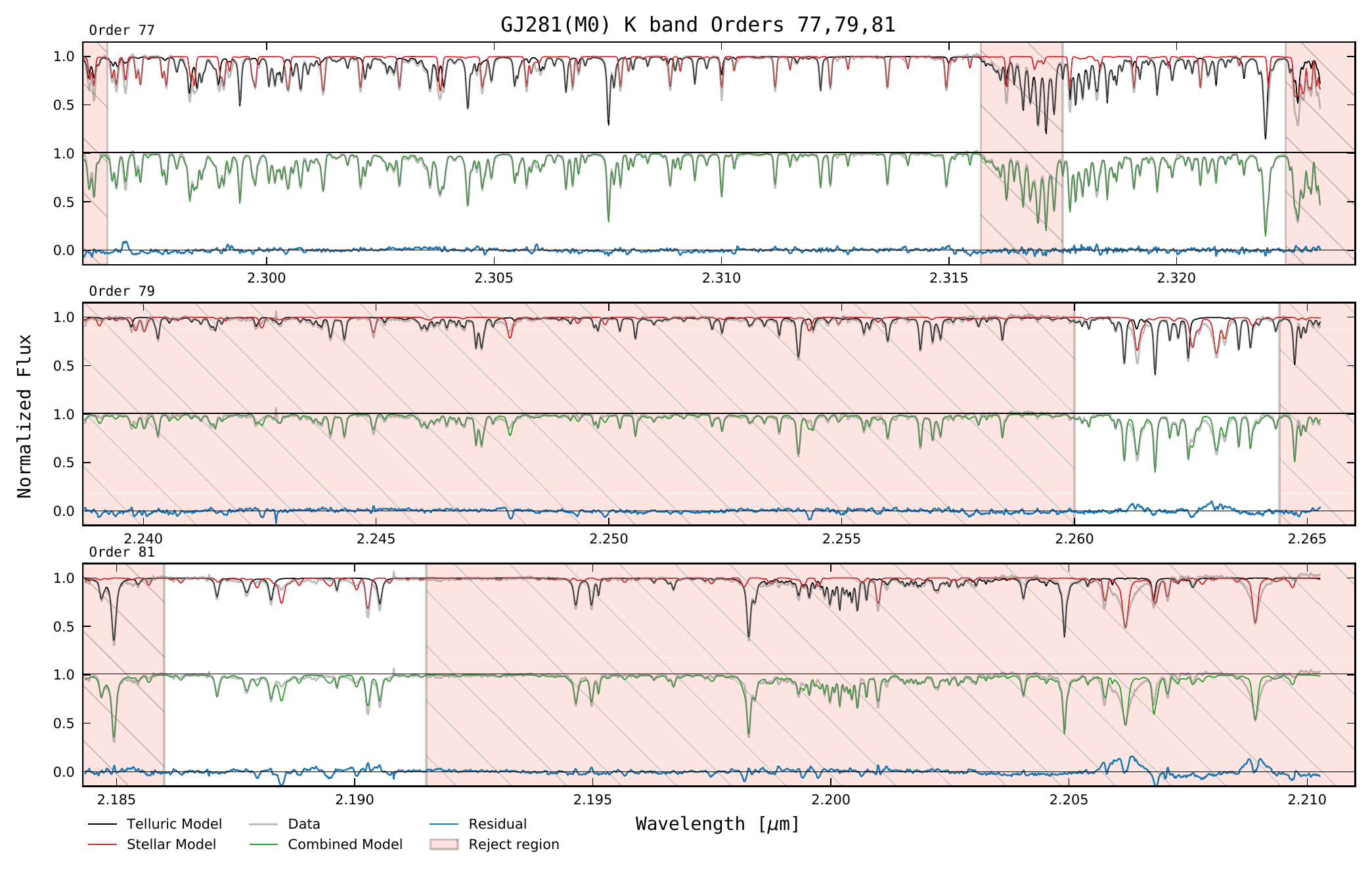}
\caption{
        Continuation of Figure~\ref{fig:GJ281_1}. 
	    }
\label{fig:GJ281_2}
\end{figure*}

\begin{figure*}[p]
\centering
\includegraphics[angle=90, width=0.8\textwidth]{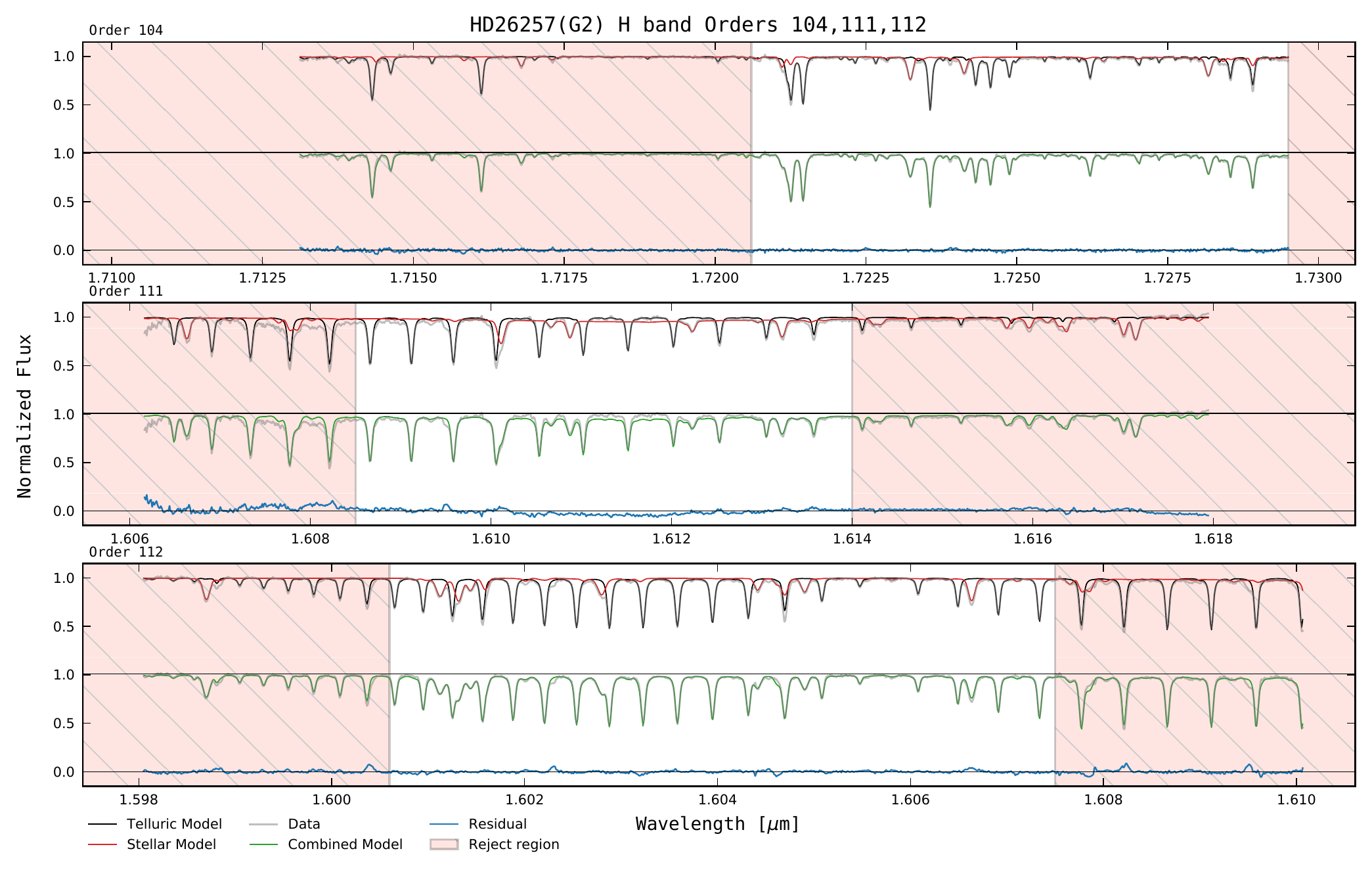}
\caption{
        Final sample of wavelength regions selected for H band RV analysis, with HD 26257 as an example star. Figure continues on next page.
	    }
\label{fig:HD26257_1}
\end{figure*}

\begin{figure*}[p]
\centering
\includegraphics[angle=90, width=0.8\textwidth]{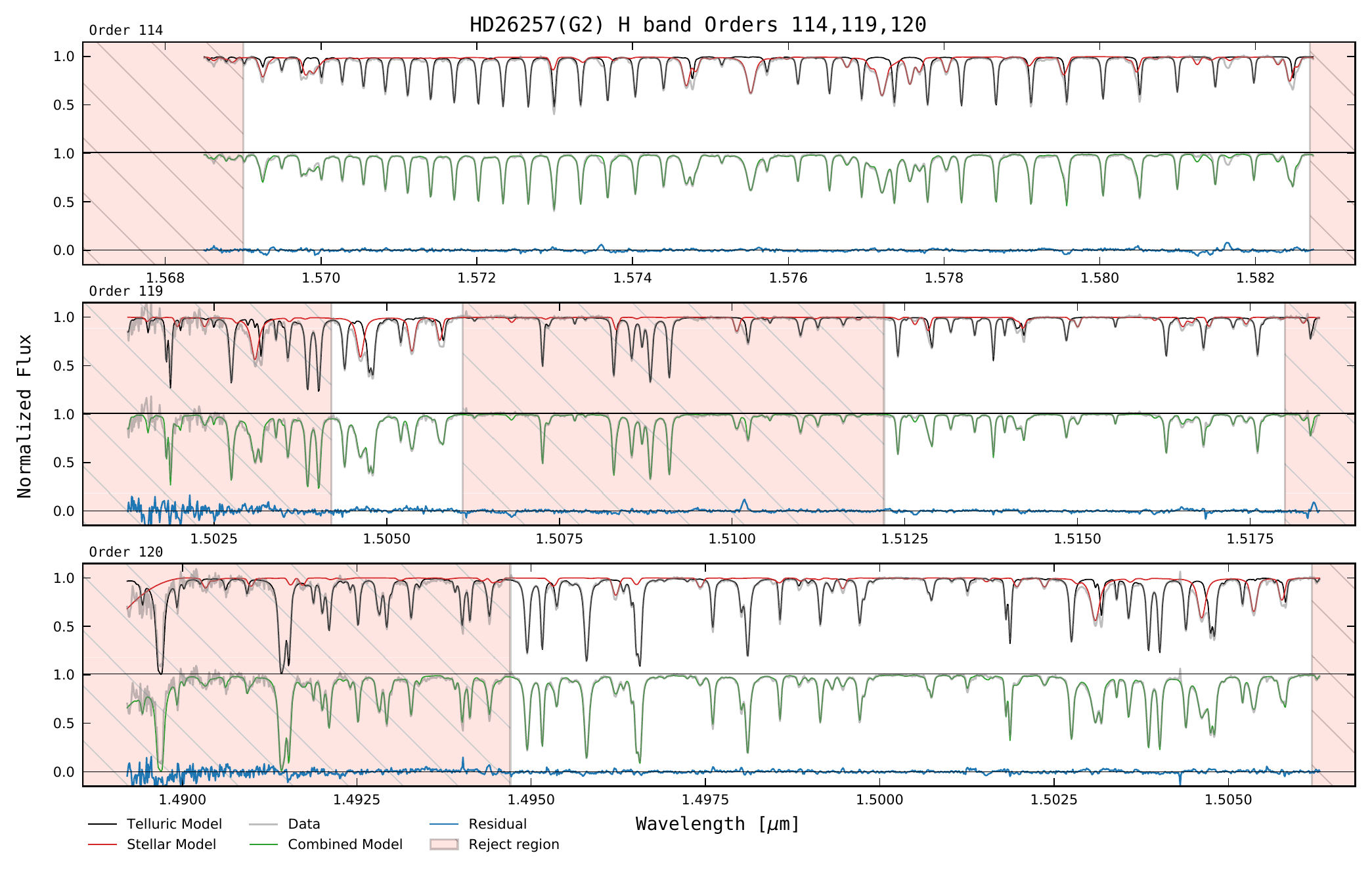}
\caption{
        Continuation of Figure~\ref{fig:HD26257_1}.
	    }
\label{fig:HD26257_2}
\end{figure*}

\newpage

\section{Observation information for Science Targets}

\newpage

\begin{longrotatetable}
\begin{deluxetable*}{crrrc r crrrc r crrrc}
\tablecaption{IGRINS Observations for RV Standards\label{tab:obs_std}
             }
\tabletypesize{\scriptsize}
\tablehead{ 
	 \colhead{UT Date}  & \colhead{S/N$^{\dagger}$} & \colhead{Nod Seq.}    & \colhead{AM} & \colhead{Facility} &&
	 \colhead{UT Date}  & \colhead{S/N$^{\dagger}$} & \colhead{Nod Seq.}    & \colhead{AM} & \colhead{Facility} &&
	 \colhead{UT Date}  & \colhead{S/N$^{\dagger}$} & \colhead{Nod Seq.}    & \colhead{AM} & \colhead{Facility} \\
	 \colhead{(yyyymmdd)} & \colhead{ }   & \colhead{ }           & \colhead{ }  & \colhead{ }         &&
	 \colhead{(yyyymmdd)} & \colhead{ }   & \colhead{ }           & \colhead{ }  & \colhead{ }         &&
	 \colhead{(yyyymmdd)} & \colhead{ }   & \colhead{ }           & \colhead{ }  & \colhead{ }         \\
     \cline{1-5} \cline{7-11} \cline{13-17} 
     \colhead{(1)} & \colhead{(2)} & \colhead{(3)} & \colhead{(4)} & \colhead{(5)} &&
     \colhead{(1)} & \colhead{(2)} & \colhead{(3)} & \colhead{(4)} & \colhead{(5)} &&
     \colhead{(1)} & \colhead{(2)} & \colhead{(3)} & \colhead{(4)} & \colhead{(5)}  
	 }
\startdata
\multicolumn{17}{c}{GJ~281 (K band)} 
\\\hline
20141122 &  55 & ABBA & 1.26 & McD && 20171005 & 109 & AB   & 1.28 & DCT && 20181019 & 119 & AB   & 1.19 & DCT\\ 
20141123 &  95 & ABBA & 1.31 & McD && 20171006 &  97 & AB   & 1.46 & DCT && 20181024 &  89 & ABBA & 1.19 & DCT\\ 
20141127 & 150 & ABBA & 1.35 & McD && 20171007 & 128 & AB   & 1.27 & DCT && 20181029 & 110 & AB   & 1.25 & DCT\\ 
20151106 & 132 & ABBA & 1.18 & McD && 20171018 & 126 & AB   & 1.20 & DCT && 20181030 &  79 & ABAB & 1.20 & DCT\\ 
20151108 & 129 & AB   & 1.25 & McD && 20171021 & 147 & AB   & 1.19 & DCT && 20181031 & 181 & AB   & 1.19 & DCT\\ 
20151111 & 135 & ABBA & 1.19 & McD && 20171028 & 162 & AB   & 1.19 & DCT && 20181125 & 124 & ABBA & 1.19 & DCT\\ 
20161219 &  72 & ABAB & 1.29 & DCT && 20171029 & 124 & AB   & 1.44 & DCT && 20181219 & 111 & AB   & 1.33 & DCT\\ 
20170106 & 114 & AB   & 1.40 & DCT && 20171119 &  89 & AB   & 1.19 & DCT && 20181220 & 161 & AB   & 1.42 & DCT\\ 
20170128 &  91 & AB   & 1.21 & DCT && 20171121 & 136 & AB   & 1.38 & DCT && 20181223 & 141 & AB   & 1.25 & DCT\\ 
20170214 & 185 & AB   & 1.27 & DCT && 20171122 & 185 & AB   & 1.40 & DCT && 20190118 & 104 & ABAB & 1.25 & DCT\\ 
20170216 &  50 & ABBA & 1.19 & DCT && 20171123 & 143 & AB   & 1.44 & DCT && 20190126 & 113 & AB   & 1.19 & DCT\\ 
20170910 & 130 & AB   & 1.73 & DCT && 20171209 & 120 & AB   & 1.58 & DCT && 20190127 & 184 & AB   & 1.20 & DCT\\ 
20170911 &  72 & ABBA & 1.75 & DCT && 20171214 & 144 & AB   & 1.35 & DCT && 20190128 &  90 & ABBA & 1.25 & DCT\\ 
20170912 & 124 & AB   & 1.79 & DCT && 20180101 & 172 & AB   & 1.21 & DCT && 20190130 & 100 & ABAB & 1.33 & DCT\\ 
20170915 & 118 & AB   & 1.60 & DCT && 20180105 & 123 & AB   & 1.19 & DCT && 20190209 & 146 & AB   & 1.31 & DCT\\ 
20170929 & 186 & AB   & 1.27 & DCT && 20180114 & 134 & AB   & 1.19 & DCT && 20190211 & 130 & AB   & 1.25 & DCT\\ 
20170930 & 178 & AB   & 1.29 & DCT && 20180123 & 156 & AB   & 1.20 & DCT && 20190304 & 135 & AB   & 1.20 & DCT\\ 
20171001 & 170 & AB   & 1.27 & DCT && 20180124 & 177 & AB   & 1.19 & DCT && 20190305 & 118 & AB   & 1.20 & DCT\\ 
20171002 & 122 & AB   & 1.33 & DCT && 20180922 & 103 & AB   & 1.46 & DCT && 20190309 &  82 & ABAB & 1.21 & DCT\\ 
20171003 & 114 & AB   & 1.37 & DCT && 20180929 & 100 & AB   & 1.75 & DCT && 20190316 & 137 & AB   & 1.25 & DCT\\ 
20171004 & 111 & AB   & 1.37 & DCT &&          &     &      &      &     &&          &     &      &      &    
\\\hline
\multicolumn{17}{c}{HD~26257 (H band)} 
\\\hline
20161110 & 120 & AB   & 1.29 & DCT && 20170911 & 113 & AB   & 1.22 & DCT && 20171011 & 176 & AB   & 1.46 & DCT\\ 
20161125 & 186 & AB   & 1.56 & DCT && 20170912 & 112 & AB   & 1.25 & DCT && 20171018 & 103 & AB   & 1.21 & DCT\\ 
20161207 & 107 & AB   & 1.39 & DCT && 20170915 & 114 & AB   & 1.22 & DCT && 20171021 & 143 & AB   & 1.24 & DCT\\ 
20161219 & 210 & AB   & 1.33 & DCT && 20170929 & 210 & AB   & 1.61 & DCT && 20171028 & 123 & AB   & 1.25 & DCT\\ 
20161220 &  66 & AB   & 1.23 & DCT && 20170930 & 189 & AB   & 1.59 & DCT && 20171029 & 121 & AB   & 1.25 & DCT\\ 
20170106 &  94 & AB   & 1.88 & DCT && 20171001 & 161 & AB   & 1.62 & DCT && 20171111 & 111 & AB   & 1.29 & DCT\\ 
20170128 & 122 & AB   & 1.27 & DCT && 20171002 & 105 & AB   & 1.21 & DCT && 20171115 & 199 & AB   & 2.13 & DCT\\ 
20170214 & 139 & BA   & 1.22 & DCT && 20171003 & 119 & AB   & 1.28 & DCT && 20171119 & 118 & AB   & 1.39 & DCT\\ 
20170215 & 187 & AB   & 1.21 & DCT && 20171004 & 111 & AB   & 1.25 & DCT && 20171209 & 119 & AB   & 1.24 & DCT\\ 
20170905 &  49 & ABBA & 1.52 & DCT && 20171005 & 115 & AB   & 1.36 & DCT && 20171213 & 108 & AB   & 1.35 & DCT\\ 
20170906 & 157 & AB   & 1.70 & DCT && 20171006 & 116 & AB   & 1.60 & DCT && 20171214 & 132 & AB   & 1.46 & DCT\\ 
20170910 & 111 & AB   & 1.26 & DCT && 20171007 & 124 & AB   & 1.23 & DCT && 20180112 & 101 & AB   & 1.27 & DCT
\enddata
\tablecomments{
    $^{\dagger}$~ Median S/N. H band from orders 104, 111, 112, 114, 119, and 120. K band from orders 74, 75, 76, and 77.
    }
\end{deluxetable*} 
\end{longrotatetable}

\startlongtable
\begin{longrotatetable}
\begin{deluxetable*}{crrrc r crrrc r crrrc}
\tablecaption{IGRINS Observations for Planet Hosts\label{tab:obs_tar}
             }
\tabletypesize{\scriptsize}
\tablehead{ 
	 \colhead{UT Date}  & \colhead{S/N$^{\dagger}$} & \colhead{Nod Seq.}    & \colhead{AM} & \colhead{Facility} &&
	 \colhead{UT Date}  & \colhead{S/N$^{\dagger}$} & \colhead{Nod Seq.}    & \colhead{AM} & \colhead{Facility} &&
	 \colhead{UT Date}  & \colhead{S/N$^{\dagger}$} & \colhead{Nod Seq.}    & \colhead{AM} & \colhead{Facility} \\
	 \colhead{(yyyymmdd\_\#)$^{\ddagger}$} & \colhead{ }   & \colhead{ }           & \colhead{ }  & \colhead{ }         &&
	 \colhead{(yyyymmdd\_\#)$^{\ddagger}$} & \colhead{ }   & \colhead{ }           & \colhead{ }  & \colhead{ }         &&
	 \colhead{(yyyymmdd\_\#)$^{\ddagger}$} & \colhead{ }   & \colhead{ }           & \colhead{ }  & \colhead{ }         \\
     \cline{1-5} \cline{7-11} \cline{13-17} 
     \colhead{(1)} & \colhead{(2)} & \colhead{(3)} & \colhead{(4)} & \colhead{(5)} &&
     \colhead{(1)} & \colhead{(2)} & \colhead{(3)} & \colhead{(4)} & \colhead{(5)} &&
     \colhead{(1)} & \colhead{(2)} & \colhead{(3)} & \colhead{(4)} & \colhead{(5)}  
	 }
\startdata
\\\hline
\multicolumn{17}{c}{HD\,189733 (K band)} 
\\\hline
20150502\_0087 & 145 & ABBA & 1.83 & McD && 20160721\_0120 & 240 & ABBA & 1.15 & McD && 20161013\_0138 & 106 & ABBA & 1.22 & DCT\\ 
20150502\_0092 &  73 & ABBAAB & 1.56 & McD && 20160721\_0124 & 241 & ABBA & 0.11 & McD && 20161013\_0142 & 103 & ABBA & 1.23 & DCT\\ 
20150502\_0104 &  47 & ABBA & 1.20 & McD && 20161013\_0038 & 110 & ABBA & 1.05 & DCT && 20161013\_0146 &  73 & ABBA & 1.25 & DCT\\ 
20150926\_0066 & 134 & ABBA & 1.12 & McD && 20161013\_0042 & 121 & ABBA & 1.05 & DCT && 20161013\_0150 &  78 & ABBA & 1.26 & DCT\\ 
20150930\_0065 & 128 & ABBA & 1.23 & McD && 20161013\_0046 & 120 & ABBA & 1.05 & DCT && 20161013\_0154 &  81 & ABBA & 1.27 & DCT\\ 
20150930\_0069 & 124 & ABBA & 1.26 & McD && 20161013\_0050 & 118 & ABBA & 1.06 & DCT && 20161025\_0043 & 123 & ABBA & 1.06 & DCT\\ 
20150930\_0073 & 122 & ABBA & 1.29 & McD && 20161013\_0054 & 121 & ABBA & 1.06 & DCT && 20161025\_0047 & 128 & ABBA & 1.06 & DCT\\ 
20150930\_0077 & 120 & ABBA & 1.32 & McD && 20161013\_0058 & 113 & ABBA & 1.06 & DCT && 20161025\_0051 & 139 & ABBA & 1.07 & DCT\\ 
20160502\_0094 & 245 & ABBA & 1.76 & McD && 20161013\_0062 & 113 & ABBA & 1.07 & DCT && 20161025\_0055 & 135 & ABBA & 1.07 & DCT\\ 
20160502\_0098 & 239 & ABBA & 1.60 & McD && 20161013\_0066 & 111 & ABBA & 1.08 & DCT && 20161025\_0059 & 118 & ABBA & 1.08 & DCT\\ 
20160502\_0106 & 262 & ABBA & 1.33 & McD && 20161013\_0070 & 103 & ABBA & 1.08 & DCT && 20161025\_0063 & 146 & ABBA & 1.09 & DCT\\ 
20160502\_0110 & 257 & ABBA & 1.25 & McD && 20161013\_0074 & 114 & ABBA & 1.08 & DCT && 20161025\_0067 & 131 & ABBA & 1.09 & DCT\\ 
20160502\_0114 & 257 & ABBA & 1.18 & McD && 20161013\_0078 & 104 & ABBA & 1.09 & DCT && 20161025\_0071 & 109 & ABBA & 1.10 & DCT\\ 
20160502\_0118 & 269 & ABBA & 1.13 & McD && 20161013\_0082 & 116 & ABBA & 1.10 & DCT && 20161025\_0075 & 139 & ABBA & 1.11 & DCT\\ 
20160502\_0126 & 234 & ABBA & 1.06 & McD && 20161013\_0086 & 113 & ABBA & 1.11 & DCT && 20161025\_0087 & 141 & ABBA & 1.16 & DCT\\ 
20160502\_0130 & 249 & ABBAAB & 1.03 & McD && 20161013\_0090 & 111 & ABBA & 1.12 & DCT && 20161025\_0091 & 146 & ABBA & 1.17 & DCT\\ 
20160721\_0049 & 258 & ABBA & 0.21 & McD && 20161013\_0094 & 116 & ABBA & 1.12 & DCT && 20161025\_0095 & 147 & ABBA & 1.18 & DCT\\ 
20160721\_0053 & 262 & ABBA & 0.75 & McD && 20161013\_0098 & 106 & ABBA & 1.13 & DCT && 20161025\_0099 & 149 & ABBA & 1.20 & DCT\\ 
20160721\_0057 & 266 & ABBA & 0.13 & McD && 20161013\_0102 &  95 & ABBA & 1.14 & DCT && 20161025\_0103 & 150 & ABBA & 1.21 & DCT\\ 
20160721\_0061 & 273 & ABBA & 0.64 & McD && 20161013\_0106 &  98 & ABBA & 1.14 & DCT && 20161025\_0107 & 143 & ABBA & 1.22 & DCT\\ 
20160721\_0065 & 275 & ABBA & 1.15 & McD && 20161013\_0110 &  95 & ABBA & 1.15 & DCT && 20161025\_0119 & 144 & ABBA & 1.29 & DCT\\ 
20160721\_0082 & 264 & ABBA & 0.53 & McD && 20161013\_0114 &  95 & ABBA & 1.16 & DCT && 20161025\_0123 & 145 & ABBA & 1.30 & DCT\\ 
20160721\_0086 & 268 & ABBA & 0.51 & McD && 20161013\_0118 & 105 & ABBA & 1.17 & DCT && 20161025\_0127 & 145 & ABBA & 1.32 & DCT\\ 
20160721\_0090 & 266 & ABBA & 0.01 & McD && 20161013\_0122 & 108 & ABBA & 1.18 & DCT && 20161025\_0131 & 149 & ABBA & 1.34 & DCT\\ 
20160721\_0094 & 261 & ABBA & 1.01 & McD && 20161013\_0126 &  93 & ABBA & 1.19 & DCT && 20161025\_0135 & 156 & ABBA & 1.36 & DCT\\ 
20160721\_0112 & 227 & ABBA & 0.03 & McD && 20161013\_0130 & 100 & ABBA & 1.20 & DCT && 20161025\_0139 & 151 & ABBA & 1.38 & DCT\\ 
20160721\_0116 & 239 & ABBA & -0.47 & McD && 20161013\_0134 & 110 & ABBA & 1.21 & DCT && 20170907\_0049 & 135 & ABBA & 1.03 & DCT
\\\hline
\multicolumn{17}{c}{$\tau$\,Boo (H band)} 
\\\hline
20150106\_0149 & 448 & ABBA & 1.46 & McD && 20150429\_0112 & 676 & ABBA & 1.03 & McD && 20160222\_0248 & 176 & ABBA & 1.16 & McD\\ 
20150106\_0153 & 291 & ABBA & 1.42 & McD && 20150429\_0116 & 690 & ABBA & 1.03 & McD && 20160222\_0252 & 204 & ABBA & 1.18 & McD\\ 
20150106\_0169 & 316 & ABBA & 1.13 & McD && 20150429\_0120 & 689 & ABBA & 1.03 & McD && 20160222\_0256 & 199 & ABBA & 1.19 & McD\\ 
20150126\_0082 & 187 & ABBA & 1.08 & McD && 20150429\_0124 & 671 & ABBA & 1.03 & McD && 20160222\_0260 & 200 & ABBA & 1.21 & McD\\ 
20150126\_0086 & 264 & ABBA & 1.07 & McD && 20150429\_0128 & 682 & ABBA & 1.03 & McD && 20160222\_0264 & 191 & ABBA & 1.22 & McD\\ 
20150126\_0090 & 233 & ABBA & 1.07 & McD && 20150429\_0136 & 689 & ABBA & 1.05 & McD && 20160224\_0116 & 298 & ABBA & 1.21 & McD\\ 
20150126\_0094 & 287 & ABBA & 1.06 & McD && 20150429\_0140 & 673 & ABBA & 1.06 & McD && 20160224\_0120 & 303 & ABBA & 1.20 & McD\\ 
20150126\_0098 & 331 & ABBA & 1.05 & McD && 20150429\_0144 & 664 & ABBA & 1.07 & McD && 20160224\_0124 & 292 & ABBA & 1.18 & McD\\ 
20150126\_0102 & 321 & ABBA & 1.05 & McD && 20150429\_0148 & 602 & ABBA & 1.09 & McD && 20160224\_0130 & 273 & ABBA & 1.17 & McD\\ 
20150126\_0130 & 276 & ABBA & 1.03 & McD && 20150429\_0152 & 589 & ABBA & 1.11 & McD && 20160224\_0134 & 292 & ABBA & 1.15 & McD\\ 
20150126\_0134 & 292 & ABBA & 1.04 & McD && 20150501\_0129 & 505 & ABBA & 1.06 & McD && 20160224\_0138 & 299 & ABBA & 1.14 & McD\\ 
20150126\_0138 & 324 & ABBA & 1.04 & McD && 20150524\_0055 & 287 & ABBA & 1.15 & McD && 20160224\_0142 & 299 & ABBA & 1.13 & McD\\ 
20150127\_0168 & 326 & ABBA & 1.11 & McD && 20150524\_0059 & 294 & ABBA & 1.13 & McD && 20160224\_0146 & 300 & ABBA & 1.12 & McD\\ 
20150127\_0172 & 328 & ABBA & 1.10 & McD && 20150524\_0063 & 304 & ABBA & 1.12 & McD && 20160224\_0150 & 294 & ABBA & 1.11 & McD\\ 
20150127\_0176 & 326 & ABBA & 1.09 & McD && 20150524\_0067 & 293 & ABBA & 1.11 & McD && 20160224\_0154 & 305 & ABBA & 1.10 & McD\\ 
20150127\_0180 & 332 & ABBA & 1.08 & McD && 20150524\_0071 & 298 & ABBA & 1.10 & McD && 20160224\_0162 & 279 & ABBA & 1.07 & McD\\ 
20150127\_0184 & 332 & ABBA & 1.07 & McD && 20150524\_0075 & 303 & ABBA & 1.09 & McD && 20160224\_0166 & 298 & ABBA & 1.06 & McD\\ 
20150127\_0188 & 312 & ABBA & 1.07 & McD && 20150524\_0079 & 292 & ABBA & 1.09 & McD && 20160224\_0170 & 296 & ABBA & 1.06 & McD\\ 
20150127\_0192 & 337 & ABBA & 1.06 & McD && 20150524\_0083 & 287 & ABBA & 1.08 & McD && 20160224\_0174 & 300 & ABBA & 1.05 & McD\\ 
20150127\_0196 & 337 & ABBA & 1.05 & McD && 20150524\_0091 & 221 & ABBA & 1.05 & McD && 20160224\_0178 & 300 & ABBA & 1.05 & McD\\ 
20150127\_0200 & 346 & ABBA & 1.05 & McD && 20150524\_0095 & 219 & ABBA & 1.05 & McD && 20160224\_0182 & 314 & ABBA & 1.04 & McD\\ 
20150127\_0204 & 339 & ABBA & 1.04 & McD && 20150524\_0099 & 219 & ABBA & 1.04 & McD && 20160224\_0186 & 297 & ABBA & 1.04 & McD\\ 
20150127\_0212 & 321 & ABBA & 1.03 & McD && 20150524\_0103 & 251 & ABBA & 1.04 & McD && 20160224\_0190 & 310 & ABBA & 1.04 & McD\\ 
20150127\_0216 & 335 & ABBA & 1.03 & McD && 20150524\_0107 & 241 & ABBA & 1.04 & McD && 20160224\_0194 & 313 & ABBA & 1.04 & McD\\ 
20150127\_0220 & 334 & ABBA & 1.03 & McD && 20150524\_0111 & 227 & ABBA & 1.03 & McD && 20160224\_0198 & 311 & ABBA & 1.03 & McD\\ 
20150127\_0224 & 324 & ABBA & 1.03 & McD && 20150524\_0115 & 232 & ABBA & 1.03 & McD && 20160224\_0202 & 303 & ABBA & 1.03 & McD\\ 
20150127\_0228 & 334 & ABBA & 1.03 & McD && 20150524\_0119 & 242 & ABBA & 1.03 & McD && 20160224\_0206 & 318 & ABBA & 1.03 & McD\\ 
20150127\_0232 & 328 & ABBA & 1.03 & McD && 20150524\_0123 & 207 & ABBA & 1.03 & McD && 20160224\_0210 & 309 & ABBA & 1.03 & McD\\ 
20150127\_0236 & 334 & ABBA & 1.03 & McD && 20150524\_0127 & 146 & ABBA & 1.03 & McD && 20160224\_0214 & 312 & ABBA & 1.03 & McD\\ 
20150127\_0240 & 326 & ABBA & 1.03 & McD && 20150524\_0131 &  49 & ABBA & 1.03 & McD && 20160224\_0218 & 308 & ABBA & 1.03 & McD\\ 
20150127\_0244 & 303 & ABBA & 1.03 & McD && 20150610\_0041 & 163 & ABBA & 1.03 & McD && 20160224\_0222 & 316 & ABBA & 1.03 & McD\\ 
20150127\_0248 & 324 & ABBA & 1.04 & McD && 20150610\_0049 & 278 & ABBA & 1.03 & McD && 20160224\_0226 & 310 & ABBA & 1.03 & McD\\ 
20150223\_0102 & 236 & ABBA & 1.27 & McD && 20150610\_0053 & 283 & ABBA & 1.03 & McD && 20160224\_0230 & 321 & ABBA & 1.03 & McD\\ 
20150223\_0106 & 243 & ABBA & 1.25 & McD && 20150610\_0059 & 266 & ABBA & 1.03 & McD && 20160224\_0234 & 311 & ABBA & 1.03 & McD\\ 
20150223\_0110 & 232 & ABBA & 1.23 & McD && 20150610\_0063 & 282 & ABBA & 1.03 & McD && 20160224\_0242 & 289 & ABBA & 1.04 & McD\\ 
20150223\_0114 & 226 & ABBA & 1.21 & McD && 20150610\_0067 & 256 & ABBA & 1.03 & McD && 20160224\_0246 & 316 & ABBA & 1.04 & McD\\ 
20150223\_0118 & 236 & ABBA & 1.19 & McD && 20150610\_0080 & 330 & ABBAAB & 1.06 & McD && 20160224\_0250 & 320 & ABBA & 1.05 & McD\\ 
20150223\_0122 & 243 & ABBA & 1.18 & McD && 20150610\_0089 & 290 & BAAB & 1.08 & McD && 20160224\_0254 & 323 & ABBA & 1.05 & McD\\ 
20150223\_0126 & 229 & ABBA & 1.16 & McD && 20150610\_0093 & 291 & BAAB & 1.09 & McD && 20160224\_0258 & 319 & ABBA & 1.06 & McD\\ 
20150223\_0131 & 239 & ABBA & 1.14 & McD && 20150610\_0097 & 333 & BAAB & 1.10 & McD && 20160224\_0262 & 319 & ABBA & 1.06 & McD\\ 
20150223\_0135 & 229 & ABBA & 1.13 & McD && 20150610\_0102 & 306 & BAAB & 1.11 & McD && 20160224\_0266 & 320 & ABBA & 1.07 & McD\\ 
20150223\_0139 & 246 & ABBA & 1.12 & McD && 20150610\_0106 & 287 & BAAB & 1.12 & McD && 20160224\_0270 & 325 & ABBA & 1.07 & McD\\ 
20150223\_0155 & 271 & ABBA & 1.05 & McD && 20150610\_0112 & 267 & BAAB & 1.13 & McD && 20160224\_0274 & 320 & ABBA & 1.08 & McD\\ 
20150223\_0159 & 266 & ABBA & 1.05 & McD && 20150610\_0116 & 287 & BAABBA & 1.15 & McD && 20160224\_0278 & 316 & ABBA & 1.09 & McD\\ 
20150223\_0163 & 261 & ABBA & 1.04 & McD && 20150630\_0057 & 284 & ABBA & 1.04 & McD && 20160224\_0282 & 328 & ABBA & 1.10 & McD\\ 
20150223\_0167 & 260 & ABBA & 1.04 & McD && 20150630\_0061 & 285 & ABBA & 1.05 & McD && 20160225\_0105 & 295 & ABBA & 1.16 & McD\\ 
20150223\_0171 & 270 & ABBA & 1.04 & McD && 20150630\_0065 & 258 & ABBA & 1.05 & McD && 20160225\_0109 & 292 & ABBA & 1.15 & McD\\ 
20150223\_0175 & 268 & ABBA & 1.03 & McD && 20150630\_0069 & 266 & ABBA & 1.06 & McD && 20160225\_0113 & 309 & ABBA & 1.14 & McD\\ 
20150223\_0179 & 263 & ABBA & 1.03 & McD && 20150630\_0073 & 201 & ABBA & 1.06 & McD && 20160225\_0117 & 307 & ABBA & 1.13 & McD\\ 
20150223\_0183 & 255 & ABBA & 1.03 & McD && 20150630\_0077 & 215 & ABBA & 1.09 & McD && 20160225\_0121 & 298 & ABBA & 1.11 & McD\\ 
20150223\_0196 & 249 & ABBA & 1.03 & McD && 20150630\_0081 & 253 & ABBA & 1.11 & McD && 20160225\_0125 & 308 & ABBA & 1.11 & McD\\ 
20150223\_0200 & 246 & ABBA & 1.03 & McD && 20150630\_0085 & 268 & ABBA & 1.12 & McD && 20160225\_0129 & 304 & ABBA & 1.10 & McD\\ 
20150223\_0204 & 249 & ABBA & 1.04 & McD && 20150630\_0089 & 259 & ABBA & 1.13 & McD && 20160225\_0133 & 316 & ABBA & 1.09 & McD\\ 
20150223\_0208 & 228 & ABBA & 1.04 & McD && 20150630\_0093 & 264 & ABBA & 1.14 & McD && 20160225\_0137 & 312 & ABBA & 1.08 & McD\\ 
20150223\_0212 & 244 & ABBA & 1.04 & McD && 20150630\_0101 & 305 & ABBA & 1.24 & McD && 20160225\_0141 & 315 & ABBA & 1.07 & McD\\ 
20150223\_0216 & 261 & ABBA & 1.05 & McD && 20150630\_0105 & 310 & ABBA & 1.26 & McD && 20160225\_0145 & 308 & ABBA & 1.07 & McD\\ 
20150306\_0108 & 431 & ABBA & 1.15 & McD && 20150630\_0109 & 319 & ABBA & 1.28 & McD && 20160225\_0153 & 323 & ABBA & 1.05 & McD\\ 
20150306\_0112 & 333 & ABBA & 1.13 & McD && 20150630\_0113 & 313 & ABBA & 1.30 & McD && 20160225\_0157 & 317 & ABBA & 1.04 & McD\\ 
20150306\_0116 & 303 & ABBA & 1.11 & McD && 20150630\_0121 & 281 & ABBA & 1.35 & McD && 20160225\_0161 & 311 & ABBA & 1.04 & McD\\ 
20150306\_0120 & 426 & ABBA & 1.09 & McD && 20150701\_0051 & 284 & ABBA & 1.13 & McD && 20160225\_0165 & 323 & ABBA & 1.03 & McD\\ 
20150306\_0124 & 408 & ABBA & 1.07 & McD && 20150702\_0051 & 305 & ABBA & 1.04 & McD && 20160225\_0171 & 323 & ABBA & 1.03 & McD\\ 
20150306\_0128 & 447 & ABBA & 1.06 & McD && 20150702\_0055 & 295 & ABBA & 1.04 & McD && 20160225\_0175 & 316 & ABBA & 1.03 & McD\\ 
20150306\_0140 & 112 & ABBA & 1.03 & McD && 20150702\_0059 & 298 & ABBA & 1.05 & McD && 20160225\_0179 & 313 & ABBA & 1.03 & McD\\ 
20150306\_0144 & 171 & ABBA & 1.04 & McD && 20150702\_0064 & 304 & ABBA & 1.06 & McD && 20160225\_0183 & 315 & ABBA & 1.03 & McD\\ 
20150306\_0148 & 236 & ABBA & 1.04 & McD && 20150703\_0059 & 283 & ABBA & 1.07 & McD && 20160225\_0187 & 324 & ABBA & 1.03 & McD\\ 
20150306\_0152 & 208 & ABBA & 1.05 & McD && 20150703\_0063 & 258 & ABBA & 1.08 & McD && 20160225\_0191 & 314 & ABBA & 1.03 & McD\\ 
20150306\_0156 & 180 & ABBA & 1.07 & McD && 20150703\_0067 & 251 & ABBA & 1.09 & McD && 20160225\_0195 & 319 & ABBA & 1.03 & McD\\ 
20150306\_0160 & 186 & ABBA & 1.08 & McD && 20150703\_0071 & 286 & ABBA & 1.10 & McD && 20160225\_0199 & 307 & ABBA & 1.03 & McD\\ 
20150306\_0173 &  80 & ABBA & 1.17 & McD && 20160202\_0156 & 275 & ABBA & 7.53 & McD && 20160225\_0203 & 311 & ABBA & 1.03 & McD\\ 
20150306\_0177 &  72 & ABBA & 1.20 & McD && 20160202\_0160 & 285 & ABBA & -1.00 & McD && 20160225\_0207 & 316 & ABBA & 1.03 & McD\\ 
20150306\_0181 &  39 & ABBA & 1.24 & McD && 20160202\_0164 & 281 & ABBA & 5.25 & McD && 20160225\_0211 & 320 & ABBA & 1.03 & McD\\ 
20150306\_0185 &  55 & ABBA & 1.29 & McD && 20160202\_0168 & 260 & ABBA & 6.73 & McD && 20160225\_0215 & 323 & ABBA & 1.04 & McD\\ 
20150306\_0189 & 111 & ABBA & 1.34 & McD && 20160202\_0172 & 280 & ABBA & 6.35 & McD && 20160225\_0219 & 317 & ABBA & 1.04 & McD\\ 
20150331\_0115 & 265 & ABBA & 1.05 & McD && 20160202\_0176 & 263 & ABBA & 6.50 & McD && 20160225\_0223 & 316 & ABBA & 1.04 & McD\\ 
20150331\_0119 & 285 & ABBA & 1.05 & McD && 20160202\_0180 & 284 & ABBA & 6.27 & McD && 20160225\_0232 & 295 & ABBA & 1.07 & McD\\ 
20150331\_0123 & 275 & ABBA & 1.04 & McD && 20160202\_0184 & 277 & ABBA & 2.53 & McD && 20160225\_0236 & 289 & ABBA & 1.07 & McD\\ 
20150331\_0127 & 246 & ABBA & 1.04 & McD && 20160202\_0188 & 291 & ABBA & 5.88 & McD && 20160225\_0240 & 288 & ABBA & 1.08 & McD\\ 
20150331\_0131 & 296 & ABBA & 1.03 & McD && 20160202\_0192 & 286 & ABBA & 5.75 & McD && 20160225\_0244 & 307 & ABBA & 1.09 & McD\\ 
20150331\_0135 & 284 & ABBA & 1.03 & McD && 20160202\_0220 & 251 & ABBA & -1.00 & McD && 20160225\_0248 & 290 & ABBA & 1.09 & McD\\ 
20150331\_0139 & 303 & ABBA & 1.03 & McD && 20160202\_0224 & 263 & ABBA & 6.03 & McD && 20160225\_0252 & 306 & ABBA & 1.10 & McD\\ 
20150331\_0143 & 301 & ABBA & 1.03 & McD && 20160202\_0228 & 246 & ABBA & 8.53 & McD && 20160225\_0256 & 302 & ABBA & 1.11 & McD\\ 
20150331\_0147 & 276 & ABBA & 1.03 & McD && 20160202\_0232 & 248 & ABBA & 8.60 & McD && 20160226\_0084 & 306 & ABBA & 1.13 & McD\\ 
20150331\_0151 & 280 & ABBA & 1.03 & McD && 20160202\_0236 & 288 & ABBA & 3.80 & McD && 20160226\_0088 & 319 & ABBA & 1.12 & McD\\ 
20150331\_0159 & 266 & ABBA & 1.03 & McD && 20160202\_0240 & 284 & ABBA & 8.60 & McD && 20160226\_0092 & 309 & ABBA & 1.11 & McD\\ 
20150331\_0163 & 252 & ABBA & 1.03 & McD && 20160202\_0244 & 289 & ABBA & 6.20 & McD && 20160226\_0096 & 317 & ABBA & 1.10 & McD\\ 
20150331\_0167 & 247 & ABBA & 1.04 & McD && 20160203\_0132 & 323 & ABBA & 1.33 & McD && 20160226\_0100 & 309 & ABBA & 1.10 & McD\\ 
20150331\_0171 & 228 & ABBA & 1.04 & McD && 20160203\_0136 & 318 & ABBA & 0.80 & McD && 20160226\_0104 & 314 & ABBA & 1.09 & McD\\ 
20150331\_0175 & 232 & ABBA & 1.04 & McD && 20160203\_0140 & 327 & ABBA & 0.80 & McD && 20160226\_0108 & 316 & ABBA & 1.08 & McD\\ 
20150331\_0179 & 235 & ABBA & 1.05 & McD && 20160203\_0144 & 319 & ABBA & -0.40 & McD && 20160226\_0112 & 308 & ABBA & 1.07 & McD\\ 
20150331\_0183 & 230 & ABBA & 1.05 & McD && 20160203\_0148 & 306 & ABBA & -0.38 & McD && 20160226\_0116 & 312 & ABBA & 1.07 & McD\\ 
20150331\_0187 & 257 & ABBA & 1.06 & McD && 20160203\_0152 & 316 & ABBA & 1.50 & McD && 20160226\_0120 & 316 & ABBA & 1.06 & McD\\ 
20150331\_0191 & 236 & ABBA & 1.06 & McD && 20160203\_0156 & 323 & ABBA & 1.58 & McD && 20160226\_0128 & 311 & ABBA & 1.04 & McD\\ 
20150331\_0195 & 257 & ABBA & 1.07 & McD && 20160203\_0160 & 311 & ABBA & 1.60 & McD && 20160226\_0132 & 316 & ABBA & 1.04 & McD\\ 
20150331\_0203 & 265 & ABBA & 1.11 & McD && 20160203\_0164 & 330 & ABBA & 1.60 & McD && 20160226\_0136 & 307 & ABBA & 1.04 & McD\\ 
20150331\_0207 & 217 & ABBA & 1.12 & McD && 20160203\_0168 & 327 & ABBA & 0.33 & McD && 20160226\_0140 & 315 & ABBA & 1.03 & McD\\ 
20150331\_0211 & 235 & ABBA & 1.13 & McD && 20160203\_0172 & 335 & ABBA & 1.70 & McD && 20160226\_0144 & 308 & ABBA & 1.03 & McD\\ 
20150331\_0215 & 218 & ABBA & 1.14 & McD && 20160203\_0176 & 333 & ABBA & -0.33 & McD && 20160226\_0148 & 316 & ABBA & 1.03 & McD\\ 
20150331\_0219 & 224 & ABBA & 1.16 & McD && 20160203\_0180 & 334 & ABBA & 1.17 & McD && 20160226\_0152 & 317 & ABBA & 1.03 & McD\\ 
20150331\_0223 & 213 & ABBA & 1.17 & McD && 20160203\_0184 & 341 & ABBA & 1.90 & McD && 20160226\_0156 & 311 & ABBA & 1.03 & McD\\ 
20150401\_0117 & 347 & ABBA & 1.05 & McD && 20160203\_0188 & 347 & ABBA & 2.25 & McD && 20160226\_0160 & 315 & ABBA & 1.03 & McD\\ 
20150401\_0121 & 335 & ABBA & 1.05 & McD && 20160203\_0192 & 344 & ABBA & 2.75 & McD && 20160226\_0164 & 317 & ABBA & 1.03 & McD\\ 
20150401\_0125 & 327 & ABBA & 1.04 & McD && 20160203\_0196 & 346 & ABBA & -0.03 & McD && 20160226\_0168 & 319 & ABBA & 1.03 & McD\\ 
20150401\_0129 & 346 & ABBA & 1.04 & McD && 20160203\_0200 & 335 & ABBA & 0.15 & McD && 20160226\_0172 & 314 & ABBA & 1.03 & McD\\ 
20150401\_0133 & 314 & ABBA & 1.03 & McD && 20160203\_0204 & 329 & ABBA & 2.55 & McD && 20160226\_0176 & 314 & ABBA & 1.03 & McD\\ 
20150401\_0137 & 339 & ABBA & 1.03 & McD && 20160222\_0064 & 242 & ABBA & 1.30 & McD && 20160226\_0180 & 310 & ABBA & 1.03 & McD\\ 
20150401\_0141 & 352 & ABBA & 1.03 & McD && 20160222\_0068 & 246 & ABBA & 1.28 & McD && 20160226\_0184 & 315 & ABBA & 1.04 & McD\\ 
20150401\_0145 & 349 & ABBA & 1.03 & McD && 20160222\_0072 & 247 & ABBA & 1.26 & McD && 20160226\_0192 & 313 & ABBA & 1.05 & McD\\ 
20150401\_0149 & 352 & ABBA & 1.03 & McD && 20160222\_0076 & 249 & ABBA & 1.24 & McD && 20160226\_0196 & 305 & ABBA & 1.06 & McD\\ 
20150401\_0153 & 340 & ABBA & 1.03 & McD && 20160222\_0080 & 249 & ABBA & 1.23 & McD && 20160226\_0200 & 306 & ABBA & 1.07 & McD\\ 
20150401\_0157 & 333 & ABBA & 1.03 & McD && 20160222\_0084 & 244 & ABBA & 1.21 & McD && 20160226\_0206 & 313 & ABBA & 1.07 & McD\\ 
20150401\_0165 & 338 & ABBA & 1.03 & McD && 20160222\_0092 & 238 & ABBA & 1.15 & McD && 20160226\_0210 & 323 & ABBA & 1.08 & McD\\ 
20150401\_0169 & 356 & ABBA & 1.03 & McD && 20160222\_0096 & 235 & ABBA & 1.14 & McD && 20160226\_0214 & 321 & ABBA & 1.09 & McD\\ 
20150401\_0173 & 355 & ABBA & 1.04 & McD && 20160222\_0100 & 239 & ABBA & 1.13 & McD && 20160226\_0218 & 319 & ABBA & 1.10 & McD\\ 
20150401\_0177 & 358 & ABBA & 1.04 & McD && 20160222\_0104 & 239 & ABBA & 1.12 & McD && 20160226\_0222 & 308 & ABBA & 1.11 & McD\\ 
20150401\_0181 & 353 & ABBA & 1.04 & McD && 20160222\_0108 & 248 & ABBA & 1.10 & McD && 20160226\_0226 & 319 & ABBA & 1.12 & McD\\ 
20150401\_0185 & 341 & ABBA & 1.05 & McD && 20160222\_0112 & 248 & ABBA & 1.09 & McD && 20160226\_0230 & 311 & ABBA & 1.13 & McD\\ 
20150401\_0189 & 346 & ABBA & 1.05 & McD && 20160222\_0120 & 225 & ABBA & 1.07 & McD && 20160226\_0234 & 316 & ABBA & 1.14 & McD\\ 
20150401\_0193 & 342 & ABBA & 1.06 & McD && 20160222\_0124 & 227 & ABBA & 1.06 & McD && 20160226\_0242 & 315 & ABBA & 1.18 & McD\\ 
20150401\_0197 & 340 & ABBA & 1.06 & McD && 20160222\_0128 & 231 & ABBA & 1.06 & McD && 20160226\_0246 & 306 & ABBA & 1.19 & McD\\ 
20150401\_0201 & 315 & ABBA & 1.07 & McD && 20160222\_0132 & 221 & ABBA & 1.05 & McD && 20160226\_0250 & 310 & ABBA & 1.21 & McD\\ 
20150401\_0205 & 355 & ABBA & 1.07 & McD && 20160222\_0136 & 233 & ABBA & 1.05 & McD && 20160226\_0254 & 315 & ABBA & 1.22 & McD\\ 
20150401\_0209 & 344 & ABBA & 1.08 & McD && 20160222\_0140 & 235 & ABBA & 1.04 & McD && 20160226\_0258 & 316 & ABBA & 1.24 & McD\\ 
20150401\_0213 & 354 & ABBA & 1.09 & McD && 20160222\_0144 & 215 & ABBA & 1.04 & McD && 20160324\_0078 & 261 & ABBA & 1.16 & McD\\ 
20150428\_0101 & 307 & ABBA & 1.23 & McD && 20160222\_0148 & 221 & ABBA & 1.04 & McD && 20160324\_0082 & 243 & ABBA & 1.15 & McD\\ 
20150428\_0105 & 292 & ABBA & 1.21 & McD && 20160222\_0152 & 220 & ABBA & 1.03 & McD && 20160324\_0086 & 246 & ABBA & 1.14 & McD\\ 
20150428\_0109 & 302 & ABBA & 1.20 & McD && 20160222\_0156 & 219 & ABBA & 1.03 & McD && 20160324\_0090 & 248 & ABBA & 1.13 & McD\\ 
20150428\_0113 & 310 & ABBA & 1.18 & McD && 20160222\_0160 & 219 & ABBA & 1.03 & McD && 20160324\_0094 & 242 & ABBA & 1.11 & McD\\ 
20150428\_0117 & 304 & ABBA & 1.17 & McD && 20160222\_0164 & 208 & ABBA & 1.03 & McD && 20160324\_0098 & 251 & ABBA & 1.10 & McD\\ 
20150428\_0121 & 306 & ABBA & 1.16 & McD && 20160222\_0168 & 198 & ABBA & 1.03 & McD && 20160324\_0107 & 263 & ABBA & 1.07 & McD\\ 
20150428\_0125 & 314 & ABBA & 1.14 & McD && 20160222\_0172 & 210 & ABBA & 1.03 & McD && 20160324\_0111 & 257 & ABBA & 1.06 & McD\\ 
20150428\_0129 & 313 & ABBA & 1.13 & McD && 20160222\_0176 & 209 & ABBA & 1.03 & McD && 20160324\_0115 & 259 & ABBA & 1.06 & McD\\ 
20150428\_0133 & 311 & ABBA & 1.12 & McD && 20160222\_0180 & 217 & ABBA & 1.03 & McD && 20160324\_0119 & 265 & ABBA & 1.05 & McD\\ 
20150428\_0137 & 303 & ABBA & 1.11 & McD && 20160222\_0184 & 205 & ABBA & 1.03 & McD && 20160324\_0123 & 261 & ABBA & 1.05 & McD\\ 
20150428\_0146 & 314 & ABBA & 1.07 & McD && 20160222\_0188 & 202 & ABBA & 1.03 & McD && 20160324\_0127 & 263 & ABBA & 1.04 & McD\\ 
20150428\_0150 & 324 & ABBA & 1.07 & McD && 20160222\_0192 & 206 & ABBA & 1.03 & McD && 20160324\_0131 & 262 & ABBA & 1.04 & McD\\ 
20150428\_0154 & 317 & ABBA & 1.06 & McD && 20160222\_0196 & 204 & ABBA & 1.03 & McD && 20160324\_0135 & 258 & ABBA & 1.04 & McD\\ 
20150428\_0158 & 326 & ABBA & 1.05 & McD && 20160222\_0200 & 196 & ABBA & 1.04 & McD && 20160324\_0139 & 257 & ABBA & 1.03 & McD\\ 
20150428\_0162 & 322 & ABBA & 1.05 & McD && 20160222\_0208 & 202 & ABBA & 1.06 & McD && 20160324\_0143 & 252 & ABBA & 1.03 & McD\\ 
20150428\_0166 & 307 & ABBA & 1.05 & McD && 20160222\_0212 & 239 & ABBA & 1.06 & McD && 20160324\_0147 & 249 & ABBA & 1.03 & McD\\ 
20150428\_0170 & 315 & ABBA & 1.04 & McD && 20160222\_0216 & 195 & ABBA & 1.07 & McD && 20160324\_0155 & 230 & ABBA & 1.03 & McD\\ 
20150428\_0174 & 317 & ABBA & 1.04 & McD && 20160222\_0220 & 204 & ABBA & 1.07 & McD && 20160324\_0159 & 240 & ABBA & 1.03 & McD\\ 
20150428\_0178 & 319 & ABBA & 1.04 & McD && 20160222\_0224 & 227 & ABBA & 1.08 & McD && 20160324\_0163 & 247 & ABBA & 1.03 & McD\\ 
20150428\_0182 & 316 & ABBA & 1.03 & McD && 20160222\_0228 & 237 & ABBA & 1.09 & McD && 20160324\_0167 & 238 & ABBA & 1.03 & McD\\ 
20150429\_0084 & 687 & ABBA & 1.12 & McD && 20160222\_0232 & 219 & ABBA & 1.09 & McD && 20160324\_0171 & 247 & ABBA & 1.04 & McD\\ 
20150429\_0088 & 691 & ABBA & 1.10 & McD && 20160222\_0236 & 236 & ABBA & 1.10 & McD && 20160324\_0175 & 248 & ABBA & 1.04 & McD\\ 
20150429\_0092 & 698 & ABBA & 1.08 & McD && 20160222\_0244 & 179 & ABBA & 1.15 & McD && 20160324\_0179 & 260 & ABBA & 1.05 & McD\\ 
20150429\_0096 & 697 & ABBA & 1.07 & McD && 20160222\_0248 & 176 & ABBA & 1.16 & McD && 20160324\_0183 & 244 & ABBA & 1.05 & McD
\enddata
\tablecomments{\\
    $^{\dagger}$~Median S/N. H band from orders 104, 111, 112, 114, 119, and 120. K band from orders 74, 75, 76, and 77.\\
    $^{\ddagger}$~``\#'' is the IGRINS fits file number, for tracking muti-RVs per night.
    }
\end{deluxetable*} 
\end{longrotatetable}

\newpage

\begin{deluxetable}{l R R c c}
\tablecaption{Telluric Standard Stars \label{tab:A0s}
             }
\tabletypesize{\scriptsize}
\tablehead{ 
	 \colhead{Designation}   & \colhead{R.A.}           & \colhead{Decl.}           & \colhead{SpTy}    & \colhead{Assoc. Targets}\\
	 \colhead{ }             & \colhead{(hh:mm:ss)}     & \colhead{($\pm$dd:mm:ss)} & \colhead{ }       & \colhead{ }    
	 }
\startdata
10 Boo & 13\colon58\colon38.921 & +21\colon41\colon46.33 & A0V & c \\
18 Ori & 05\colon16\colon04.135 & +11\colon20\colon28.88 & A0V & ab \\
29 Vul & 20\colon38\colon31.329 & +21\colon12\colon04.38 & A0V & d \\
30 Mon & 08\colon25\colon39.632 & -03\colon54\colon23.12 & A0V & a \\
33 LMi & 10\colon31\colon51.375 & +32\colon22\colon46.41 & A0V & a \\
5 Vul & 19\colon26\colon13.246 & +20\colon05\colon51.84 & A0V & d \\
HD 121880 & 13\colon57\colon52.121 & +16\colon12\colon07.52 & A0V & c \\
HD 184787 & 19\colon34\colon18.191 & +41\colon55\colon37.99 & A0V & d \\
HD 31743 & 04\colon57\colon44.641 & -13\colon42\colon17.93 & A0V & a \\
HD 53205 & 07\colon04\colon20.175 & +01\colon29\colon18.57 & B9IV/V & ab \\
HD 63714 & 07\colon50\colon13.993 & -01\colon27\colon32.12 & A0V & a \\
HD 65158 & 07\colon57\colon25.860 & +00\colon38\colon09.15 & A0V & a \\
HR 1039 & 03\colon27\colon18.676 & +12\colon44\colon07.03 & A0V & ab \\
HR 1137 & 03\colon44\colon28.204 & +20\colon55\colon43.45 & A0V & b \\
HR 1234 & 04\colon01\colon14.832 & +36\colon59\colon22.10 & A0V & a \\
HR 1237 & 04\colon00\colon36.888 & +17\colon17\colon47.97 & A0V & ab \\
HR 1367 & 04\colon20\colon39.013 & -20\colon38\colon22.63 & A0V & a \\
HR 1482 & 04\colon41\colon24.128 & +48\colon18\colon03.16 & A0V & ab \\
HR 1558 & 04\colon54\colon51.243 & +44\colon03\colon39.11 & A0V & b \\
HR 1578 & 04\colon55\colon58.352 & +05\colon23\colon56.49 & A0V & b \\
HR 1724 & 05\colon16\colon41.039 & +01\colon56\colon50.38 & A0V & b \\
HR 2133 & 06\colon03\colon24.769 & +11\colon40\colon51.84 & A0V & a \\
HR 2315 & 06\colon25\colon46.528 & +02\colon16\colon18.04 & A0V & a \\
HR 2584 & 06\colon55\colon34.618 & +08\colon19\colon27.43 & A0V & a \\
HR 2893 & 07\colon34\colon04.962 & +10\colon34\colon05.07 & A0V & a \\
HR 7734 & 20\colon14\colon04.883 & +36\colon36\colon17.56 & A0V & d \\
HR 945 & 03\colon10\colon08.839 & +27\colon49\colon11.16 & A0V & b \\
k Tau & 04\colon58\colon09.392 & +25\colon03\colon01.47 & A0V & ab \\
NSV12680 & 20\colon00\colon15.543 & +29\colon55\colon14.25 & A0V & d
\enddata
\tablecomments{ $\rm^{a}$GJ\,281, $\rm^{b}$HD\,26257, $\rm^{c}\tau$Boo\,A, $\rm^{d}$HD\,189733}
\end{deluxetable}

\bibliography{igrinsrv}
\bibliographystyle{aasjournal}


\end{document}